\documentclass[traditabstract]{aa}
\usepackage{epsfig}    
\usepackage{lscape}
\usepackage{natbib}
\bibpunct{(}{)}{;}{a}{}{,}    
\usepackage{graphics}
\usepackage{graphicx,graphics}
\usepackage{color} 
\usepackage{times}
\usepackage{amsmath}
\usepackage{amssymb}
\begin{document}

\title {MaNGA galaxies with off-centered spots of enhanced gas velocity dispersion}

\author {
  L.~S.~Pilyugin\inst{\ref{MAO}}             \and 
  B.~Cedr\'{e}s\inst{\ref{IAC},\ref{ULL}}    \and 
  I.~A.~Zinchenko\inst{\ref{LMU},\ref{MAO}}  \and
  A. M.~P\'{e}rez Garcia\inst{\ref{CAM},\ref{AAp}}    \and 
  M.~A.~Lara-L\'{o}pez\inst{\ref{AOP}}       \and
  J.~Nadolny\inst{\ref{IAC},\ref{ULL}}       \and 
  Y.~A.~Nefedyev\inst{\ref{KGU}}             \and
  M.~Gonz\'{a}lez-Otero\inst{\ref{IAC},\ref{ULL}}  \and 
  J.~M.~V\'{i}lchez\inst{\ref{IAA}}          \and
  S.~Duarte~Puertas\inst{\ref{Can}}          \and
  R.~Navarro Martinez\inst{\ref{IAC},\ref{ULL}} 
         }
\institute{Main Astronomical Observatory, National Academy of Sciences of Ukraine, 27 Akademika Zabolotnoho St, 03680, Kiev, Ukraine \label{MAO} \and
Instituto de Astrof\'{i}sica de Canarias (IAC), E-38200 La Laguna, Tenerife, Spain \label{IAC} \and  
Departamento de Astrof\'{i}sica, Universidad de la Laguna (ULL), 38205 La Laguna, Tenerife, Spain\label{ULL} \and 
Faculty of Physics, Ludwig-Maximilians-Universit\"{a}t, Scheinerstr. 1, 81679 Munich, Germany \label{LMU}  \and
Centro de Astrobiolog\'{i}a (CSIC/INTA), 28692 ESAC Campus, Villanueva de la Ca\~{n}ada, Madrid, Spain \label{CAM} \and
Asociaci\'{o}n Astrof\'{i}sica para la Promoci\'{o}n de la Investigaci\'{o}n, Instrumentaci\'{o}n y su Desarrollo, ASPID, 38205 La Laguna, Tenerife, Spain \label{AAp} \and
Armagh Observatory and Planetarium, College Hill, Armagh BT61 9DG, Northern Ireland, UK \label{AOP} \and 
Kazan Federal University, 18 Kremlyovskaya St., 420008, Kazan, Russian Federation \label{KGU} \and 
Instituto de Astrof\'{i}sica de Andaluc\'{i}a, CSIC, Apdo 3004, 18080 Granada, Spain \label{IAA} \and
D\'epartement de Physique, de G\'enie Physique et d'Optique, Universit\'e Laval, and Centre de Recherche en Astrophysique du Qu\'ebec (CRAQ), Qu\'ebec, QC, G1V 0A6,
Canada \label{Can}
}

\abstract{Off-centered spots of the enhanced gas velocity dispersion, $\sigma$, are revealed in some galaxies from the  Mapping Nearby Galaxies at Apache Point
Observatory survey (MaNGA).  Aiming to clarify the origin of the spots of  enhanced $\sigma$, we examine the distributions of the surface brightness,
the line-of-sight velocity, the oxygen abundance, the gas velocity dispersion, and the Baldwin-Phillips-Terlevich (BPT) spaxel classification in seven galaxies.
We find that the enhanced $\sigma$ spots in six galaxies can be attributed to a (minor) interaction with a satellite. Three galaxies in our sample have a very close
satellite (the separation in the sky plane is comparable to the optical radius of the galaxy). The spots of enhanced $\sigma$ in those galaxies are located at the
edge of the galaxy close to the satellite. The spots of  enhanced $\sigma$ in three other galaxies are related to bright spots in the photometric $B$ band within
the galaxy, which can be due to the projection of a satellite in the line of sight of the galaxy. The oxygen abundances in the spots in these three galaxies are
reduced. This suggests that the low-metallicity gas from the satellite is mixed with the interstellar medium of the disk, that is, the gas exchange between the
galaxy and its satellite takes place. The spectra of the spaxels within a spot are usually  H\,{\sc ii}-region-like, suggesting that the interaction (gas infall)
in those galaxies does not result in appreciable shocks. In contrast, the spot of the enhanced $\sigma$ in the galaxy M-8716-12703 is associated with an off-centered
active galactic nucleus-like (AGN-like) radiation distribution. One can suggest that the spot of the enhanced $\sigma$ in the M-8716-12703 galaxy is different in origin,
or that the characteristics of gas infall in this case differs from that in other galaxies. 
}

\keywords{galaxies: ISM -- galaxies: kinematics and dynamics -- galaxies: structure}

\titlerunning{off-centered spots of enhanced $\sigma$ in galaxies}
\authorrunning{Pilyugin et al.}
\maketitle

\section{Introduction}

The gas velocity dispersion of nearby disk galaxies traced by ionized gas is higher than the expected dispersion due to thermal motion alone
\citep{Epinat2010,Varidel2020,Bacchini2020}, and there is a trend of increasing  gas velocity dispersion with increasing  redshift
\citep{Johnson2018,Ubler2019,Hung2019}. Although several mechanisms for the enhancement of the gas velocity dispersion in disk galaxies (including feedback
from star formation activity and a galaxy interaction and merger or intergalactic gas acretion) have been considered; the drivers responsible for the enhancement
of the gas velocity dispersion are still debated \citep[][and references therein]{Stilp2013,Zhou2017,Hung2019,Kohandel2020,Hunter2021}.

Numerical simulations predict that the enhancement of the gas velocity dispersion in the disk can be caused by feedback from star formation activity,
including supernovae and winds from stars \citep[e.g.,][]{Dib2006}. The validity of the predicted relationship between the velocity dispersion of the gas and the
star formation rate (or supernova rate) has been tested through the comparison with observations. However, the available observations have not decisively identified
the main driver of the gas velocity dispersion. Indeed, \citet{Lehnert2013} and \citet{Bacchini2020}  conclude that supernovae alone can sustain gas turbulence
in galaxies and that there is essentially no need for any further source of energy. On the contrary, \citet{Krumholz2018} and \citet{Varidel2020}  found that the radial
gas transport should be added to star formation feedback in order to explain the observed gas velocity dispersion in galaxies. 

The enhancement of the gas velocity dispersion in the disk can be caused by interaction and merger or accretion \citep{Klessen2010,Bournaud2011}.  
Simulations of interactions and mergers of galaxies are the topic of many works  \citep[][among many others]{Walker1996,Naab2003,
Bournaud2004,Springel2005,Robertson2006,Governato2007,Lotz2008,Hopkins2009a,Hopkins2009b,Zinchenko2015,RodriguezGomez2017}.
It is established that the observed properties of the interaction or merger remnants depend on the characteristics of the progenitors,
the geometry of the collision, and the merger stage. The properties of the merger remnant depend strongly on the mass ratio of the progenitors.
Major galaxy mergers with mass ratios in the range 1:1 -- 3:1 lead to the formation of boxy or disky elliptical galaxies, mergers with mass ratios
in the intermediate range 4:1 -- 10:1 result in peculiar galaxies with disk-type morphologies, but kinematics closer to that of elliptical systems. On the other hand, 
minor mergers with mass ratios below 10:1 result in disturbed spiral galaxies \citep{Bournaud2004}. The difference between minor merger and
accretion is blured \citep{Hopkins2009b}.
The properties of the merger remnant are also determined by the gas fraction of the progenitors.  When the gas fraction of the
progenitors is low then the remnants structurally and kinematically resemble elliptical galaxies.  If the progenitor galaxies are gas-rich
then a prominent preexisting disk can survive, that is, both major and minor mergers can produce a disk-dominated galaxy.
\citet{RodriguezGomez2017} considered the influence of mergers in the galaxy morphology using the Illustris simulations. They found that
mergers play a dominant role in shaping the morphology of massive galaxies, while mergers do not seem to play any significant role in
determining the morphology of galaxies with masses below $\sim$10$^{11}M_{\odot}$.

This is the third paper of series devoted to the investigation of the distribution of the gas velocity dispersion $\sigma$ in
star-forming (SF) galaxies from the Mapping Nearby Galaxies at Apache Point  Observatory (MaNGA)  survey \citep{Bundy2015,Albareti2017}.  
In the first paper, we considered the circumnuclear regions of 161 galaxies from the Sloan Digital Sky Survey Data Release 15 (SDSS DR15) MaNGA  survey
\citep{Pilyugin2020b}. The spaxel spectra were classified as  active galactic nucleus-like (AGN-like), H\,{\sc ii}-region-like (or SF-like), and
intermediate (INT) spectra according to their positions in the Baldwin-Phillips-Terlevich \citep[BPT, ][]{Baldwin1981} diagnostic diagram.
There are four configurations in the type of radiation of the circumnuclear regions:
1) AGN, the innermost region is AGN-like radiation and is surrounded by a ring of  radiation of the intermediate type;
2) INT, the central area of radiation is of the intermediate type; 
3) SF+INT, the inner region is H\,{\sc ii}-like radiation and is surrounded by a ring of radiation of the intermediate type; and 
4) SF, the central area is H\,{\sc ii}-like radiation only.
We found that the AGN-like and INT radiation in the circumnuclear
region is accompanied by an enhancement in the gas velocity dispersion $\sigma$, in the sense that the  gas velocity dispersion in
a galaxy decreases with galactocentric distance up to some radius and remains approximately constant beyond this radius.
The radius of the area of the AGN-like and INT radiation (radius of influence of the AGN on the radiation) 
is similar to the radius of the area with enhanced gas velocity dispersion, and the central gas velocity dispersion
$\sigma_{c}$ correlates with the luminosity of the AGN+INT area. 
This suggests that the enhancement of the gas velocity dispersion at the center of a galaxy can be attributed to the AGN activity.

In the second paper, the distribution of the gas velocity dispersion $\sigma$ across the images of 1146 MaNGA galaxies beyond the
centers is analyzed  \citep{Pilyugin2021}. We find that there are two types of distribution of the gas velocity dispersion  across
the images of galaxies: $(i)$ the distributions of 909 galaxies show a radial symmetry with or without a $\sigma$ enhancement
at the center (R distribution) and $(ii)$ distributions with a band of enhanced $\sigma$ along the minor axis in the images of 159
galaxies with or without a $\sigma$ enhancement at the center (B distribution). The $\sigma$ distribution across the images of
78 galaxies cannot be reliable classified. We find that the median value of the gas velocity dispersion $\sigma_{m}$ in galaxies with
B distribution is higher by around 5 km/s, on average, than that of galaxies with R distribution. The optical radius $R_{25}$ of galaxies
with B distribution is lower by around 0.1 dex, on average,  than that of galaxies with similar masses with R distribution. Thus the
properties of a galaxy are related to the type of  distribution of the gas velocity dispersion $\sigma$ across its image. 
This suggests that the presence of the band of the enhanced gas velocity dispersion can be an indicator of a specific evolution
(or a specific stage in the evolution) of a galaxy.   

We find that there are off-centered spots of enhanced gas velocity dispersion $\sigma$ in some MaNGA galaxies. The global distribution of the gas velocity
dispersion $\sigma$ across the image of the galaxy is more or less regular (with or without the $\sigma$ enhancement at the center and with or without
the band of enhanced $\sigma$ along the minor axis), and values of the gas velocity dispersion in the spot show an evident deviation from the global
distribution. To examine the  origin of the enhanced $\sigma$ spots, we produce and analyze  maps of the surface brightness,  line-of-sight velocity,
 oxygen abundance,  gas velocity dispersion, and  BPT  spaxel classification in seven MaNGA galaxies that show off-centered spots of  enhanced
gas velocity dispersion. The parameters of emission lines in each spaxel spectrum are estimated using a single Gaussian fit to the line profiles. The sigma of
the best-fit Gaussian of the  H$\alpha$ line  is converted into the gas velocity dispersion $\sigma$.

This paper is organized as follows. The data are described in Section 2. In Section 3, the distributions of the gas velocity dispersion and
other characteristics in the MaNGA galaxies are constructed and the comments on individual galaxies are given. In Section 4 a discussion is given,
and Section 5 provides a brief summary.

\section{Data}

The publicly available spectroscopic observations from the Sloan Digital Sky Survey Data Release 15 (SDSS DR15) MaNGA  survey \citep{Bundy2015,Albareti2017}
are at the base of this investigation. The same data was also used in our previous studies \citep{Pilyugin2020b,Pilyugin2021}, and the data reduction is
described in \citet{Zinchenko2016} and \citet{Sakhibov2018}. Briefly, we used LOGCUBE datacubes, which have logarithmic wavelength sampling.
The stellar radiation contribution is approximated using the public version of the STARLIGHT code
\citep{CidFernandes2005,Mateus2006,Asari2007}, which was adapted for execution in the NorduGrid Advanced Resource Connector
(ARC)\footnote{http://www.nordugrid.org/} environment of the Ukrainian National Grid. We use a sample of stellar population (SSP) spectra
from the evolutionary synthesis models of \citet{Bruzual2003}. The correction for  reddening is performed using the curve from \citet{Cardelli1989}
with $R_V = 3.1$. The gaseous spectrum is obtained by  subtracting the derived stellar radiation contribution from the observed spectrum. 

The  emission line profiles in each spaxel spectrum were fitted by single Gaussians 
using our emission line fitting code ELF3D. The code is based on the iminuit library. Since the applied fitting scheme is sensitive
to the choice of the initial  parameters, we implemented a Monte Carlo (MC) approach for choosing the initial parameters of the fit, which significantly
increased the robustness of 
the fitting. ELF3D allows to fit lines independently or in groups where central wavelengths and/or widths of profiles can be tied.
In this work we tied the center and width of the following lines doublets [O\,{\sc ii}]$\lambda\lambda$3727,3729, [O\,{\sc iii}]$\lambda\lambda$4959,5007,
and [N\,{\sc ii}]$\lambda\lambda$6548,6584. Also, the lines center of the [N\,{\sc ii}]$\lambda\lambda$6548,6584 doublet have been tied to H$\alpha$ 
as they have been fitted in the same group.

The estimated emission line parameters 
are the central wavelength $\lambda_{0}$, the sigma $\sigma$, and the flux $F$. For each spectrum, the [O\,{\sc ii}]$\lambda\lambda$3727,3729, H$\beta$,
[O\,{\sc iii}]$\lambda$5007, H$\alpha$, and [N\,{\sc ii}]$\lambda$6584 emission lines were measured. In the construction of each particular map, only those
spaxel spectra where the concerned lines were measured with a signal-to-noise ratio S/N $> 3$ were used. In some spectra, the strong H$\alpha$ line
is measured with S/N $>$ 3 while the weak lines (e.g. [N\,{\sc ii}]$\lambda$6584 line) are measured with S/N $<$ 3. Therefore the different maps for a given
galaxy can contain different numbers of the spaxels. The central wavelength of the H$\alpha$ line was converted into the line-of-sight velocity $V_{los}$, and  
the sigma of the best-fit Gaussian $\sigma_{{\rm H}\alpha}$ was converted into the observed gas velocity dispersion $\sigma_{obs}$. Since the values of the observed
(non-corrected-for instrumental profile) gas velocity dispersion are used throughout the paper, we use the notation $\sigma$ instead of  $\sigma_{obs}$. 
  
The interstellar reddening is estimated through the comparison between the measured and the theoretical H$\alpha$/H$\beta$ ratios using the reddening law from
\citet{Cardelli1989} for $R_{V}$ = 3.1.  It was adopted  $C_{{\rm H}{\beta}} = 0.47A_{V}$ \citep{Lee2005}.
We classify the excitation of the spaxel spectrum using its position on the
 standard diagnostic diagram of the [N\,{\sc ii}]$\lambda$6584/H$\alpha$ versus the [O\,{\sc iii}]$\lambda$5007/H$\beta$ line ratios suggested by
\citet{Baldwin1981}, which is known as the BPT classification diagram.
As in our previous studies \citep{Zinchenko2019,Pilyugin2020a,Pilyugin2020b,Pilyugin2021},
the spectra located to the left (below) the demarcation line from \citet{Kauffmann2003} are referred to as the SF-like or
H\,{\sc ii} region-like spectra, those located to the right (above) the demarcation line from \citet{Kewley2001} are referred to as the AGN-like spectra,
and the spectra located between those demarcation lines are  the intermediate (INT) spectra.

The geometrical parameters of the galaxy needed to determine  the galactocentric distances of the spaxels, are
derived from the analysis of the observed velocity field in the standard way assuming that a galaxy is a symmetrically rotating disk
\citep[e.g.,][]{Warner1973,Begeman1989,deBlok2008,Oh2018}. The position of the kinematic center of the galaxy, the position angle of the major kinematic axis
and the kinematic inclination angle are determined from the measured line-of-sight gas velocities (obtained from the H$\alpha$ line) in the same way
as in our previous papers \citep{Pilyugin2019,Pilyugin2020a,Zinchenko2019}. Briefly, the observed line-of-sight velocities recorded on a set of pixel coordinates
(1 pixel = 0.5 arcsec for MaNGA galaxies) are related to the kinematical parameters of the galaxy and its rotation curve. We divide the deprojected galaxy plane 
into rings with a width of one pixel. It is assumed that the rotation velocity is the same for all the spaxels within the ring. We adopt that the position angle
of the major kinematic axis and the galaxy inclination angle are the same for all the rings. The coordinates of the rotation center of the galaxy,
the position angle of the major kinematic axis, the galaxy inclination angle, and the rotation curve are determined through the best fit of the line-of-sight velocity
field $V_{los}$. We follow two steps for deriving  the rotation curve and the geometrical parameters of each galaxy.  In the first step, the values of the parameters
are obtained for all the spaxels with measured $V_{los}$. In the second step, we use an iterative procedure to determine the rotation curve  and the geometrical parameters.
At each step, points with large deviations from the rotation curve determined in the previous step are rejected, and new values of the
geometrical parameters and the rotation curve are derived.
The difference between the values of the spaxel velocity obtained from the measured wavelengths of the H$\beta$ and H$\alpha$ lines can be considered as
some kind of estimation of the error in the spaxel line-of-sight velocity measurements in the MaNGA spectra. The mean value of the differences  between
the measured  H$\beta$ and H$\alpha$ velocities in 46,350  spaxels in the MaNGA galaxies is $\sim$7 km/s \citep{Pilyugin2019}.
The points with deviations larger than 21 km/s (three sigma) from the rotation curve determined in the previous step are rejected. 
The iteration is stopped when the absolute values of the difference of coordinates of the center obtained
in successive steps is less than 0.1 pixels, the difference of position angle of the major axis and the inclination angle is smaller than 0.1$\degr$, and the rotation
curves agree within 1 km/s (at each radius).  A detailed discussion of the determination of coordinates of the rotation center, the inclination angle, the position angle
of the major kinematic axis, and the rotation curve can be found in  \citet{Pilyugin2019}

The surface brightness in the SDSS $g$ and $r$ bands for each spaxel was obtained from broadband SDSS images created from the data cube. The measured magnitudes are
converted to $B$-band magnitudes and  corrected for Galactic foreground extinction using the recalibrated $A_V$ values from \citet{Schlafly2011} as reported in the
NASA/IPAC Extragalactic Database ({\sc ned})\footnote{The NASA/IPAC Extragalactic Database ({\sc ned}) is operated by the Jet Propulsion Laboratory, California Institute
of Technology, under contract with the National Aeronautics and Space Administration.  {\tt http://ned.ipac.caltech.edu/}}.
The observed surface-brightness profile is obtained using  the geometrical parameters of the galaxy derived from the observed velocity field analysis (the position of
the kinematic center of the galaxy, the position angle of the major kinematic axis and the kinematic inclination angle). The observed surface-brightness profile within
a galaxy was fitted by a broken exponential profile for the disk and by a general S\'{e}rsic profile for the bulge \citep{Pilyugin2014,Pilyugin2017,Pilyugin2018}.
The optical radius of the galaxy $R_{25}$ was estimated using the obtained fit.

The distances to the galaxies were taken from {\sc ned}. The {\sc ned} distances use flow corrections for Virgo, the Great Attractor, and Shapley Supercluster infall
(adopting a cosmological model with $H_{0} = 73$ km/s/Mpc, $\Omega_{m} = 0.27$, and $\Omega_{\Lambda} = 0.73$). We have chosen the spectroscopic $M_{sp}$ masses of the SDSS
and BOSS \citep[BOSS stands for the Baryon Oscillation Spectroscopic Survey in SDSS-III, see][]{Dawson2013}.  The spectroscopic masses are taken from the
table {\sc stellarMassPCAWiscBC03}, and were determined using the Wisconsin method \citep{Chen2012} with the stellar population synthesis models from \citet{Bruzual2003}.

\begin{table*}
\caption[]{\label{table:sample}
Properties of our sample of MaNGA galaxies with the off-centered spots of the enhanced gas velocity dispersion. }
\begin{center}
\begin{tabular}{ccccccccccc} \hline \hline
MaNGA                 &
other                 &
RA                    &
DEC                   &
$i$                   &
log$M_{sp}$            &
R$_{25}$               &
d                     &
R$_{spot}$             & 
BPT$_{center}$          &           
BPT$_{spot}$           \\
number                &
name                  &
($\degr$)             &
($\degr$)             &
($\degr$)             &
($M_{\odot}$)          &           
(Kpc)                 &           
(Mpc)                 &
($R_{25}$)             &
                      &           
                     \\   \hline
 7977 -- 09102       &            &  332.830664 &   12.184717 &   41.9 &  10.935 &  12.8 &  251.4 &  1.0 &  1 &  1 \\
 8080 -- 09101       &            &   47.340029 &   -0.407944 &   57.3 &   9.881 &  10.5 &  180.3 &  1.0 &  0 &  0 \\                      
 8244 -- 09101       & PGC~25036  &  133.787581 &   52.470442 &   27.4 &  11.201 &  18.7 &  241.6 &  0.8 &  0 &  0 \\ 
 8568 -- 12703       &            &  156.810365 &   38.204502 &   26.6 &  10.469 &  14.5 &  221.2 &  0.8 &  0 &  0 \\ 
 8716 -- 12703       &            &  122.411182 &   54.552173 &   43.1 &   9.770 &   9.3 &  174.5 &  0.7 &  0 &  2 \\ 
 8984 -- 12705       & NGC~5251   &  204.353484 &   27.419218 &   36.5 &  11.292 &  20.3 &  158.3 &  0.6 &  2 &  0 \\ 
 9031 -- 06104       &            &  240.585568 &   44.470175 &   41.7 &  10.019 &   8.5 &  185.1 &  0.9 &  0 &  0 \\ 
 \hline
\end{tabular}\\
\end{center}
\begin{flushleft}
{\bf Notes.} The columns show
the MaNGA number,
the other name,
the right ascension (RA) and the declination (DEC) (J2000.0), 
the kinematic inclination angle $i$,
the spectroscopic stellar mass $M_{sp}$ in solar masses,
the optical radius $R_{25}$ in kpc, 
the galaxy distance $d$ in Mpc,
the galactocentric distance (normalized to the optical radius $R_{25}$) of the spot position, 
the BPT type of the radiation at the center of the galaxy (0 -- SF, 1 -- intermediate, 2 -- AGN),
the BPT type of the radiation in the spot of the enhanced gas velocity dispersion.
\end{flushleft}
\end{table*}

The oxygen abundances were obtained through the three-dimensional $R$ calibration \citep{Pilyugin2016,Pilyugin2018}. The [O\,{\sc ii}]$\lambda\lambda$3727,3729,
H$\beta$, [O\,{\sc iii}]$\lambda\lambda$4959,5007, and [N\,{\sc ii}]$\lambda\lambda$6548,6584 emission lines are used for oxygen abundance determinations. 
The flux in the [O\,{\sc iii}]$\lambda\lambda$4959,5007 lines is estimated as [O\,{\sc iii}]$\lambda\lambda$4959,5007 = 1.3[O\,{\sc iii}]$\lambda$5007
and the flux in the [N\,{\sc ii}]$\lambda\lambda$6548,6584 lines is estimated as [N\,{\sc ii}]$\lambda,\lambda$6548,6584 = 1.3[N\,{\sc ii}]$\lambda$6584  \citep{Storey2000}.
The $R$-calibration produces abundances compatible to the $T_{e}$-based abundance scale and is valid over the whole metallicity scale of H\,{\sc ii} regions. 

We carried out an extensive search for  galaxies with  off-centered spots of  enhanced $\sigma$ by performing visual examination of the $\sigma$ maps of about three
thousands galaxies from the SDSS DR15 MaNGA survey but we do not pretend that our list is exhaustive. The visual selection of  spots with enhanced gas velocity dispersion
is somewhat arbitrary. Our preliminary list includes 14 MaNGA galaxies with off-centered spots of  enhanced gas velocity dispersion $\sigma$. There is a variation
in the values of  $\sigma$ across the image of each galaxy. The variation of $\sigma$ can form a global structure: the $\sigma$ distribution can show a band of
enhanced $\sigma$ along the minor axis in the image \citep{Pilyugin2021}, and the value of the gas velocity dispersion is usually enhanced at the center of the galaxy
\citep{Pilyugin2020b}. There is a chaotic variation in the $\sigma$ in addition to the global variation. The amplitude of the chaotic variation of the $\sigma$ is more
or less similar over the whole image of the galaxy. In some MaNGA galaxies, however, we identify off-centered spots of  enhanced gas velocity dispersion where the
$\sigma$ enhancement in the spot is significantly higher than the amplitude of the chaotic variation.

Seven galaxies of the preliminary list were excluded from futher consideration for the following reasons. In three galaxies we find that the emission lines
in the spectra of the spaxels within the spots  are double or double-peaked
(M-8553-09102, M-9029-12705, and M-9049-12701).
Those galaxies were excluded from our current sample since the large widths of the emission line profiles correspond to two Gaussians, and not one.
We find that the validity of some line measurements in the spots of four galaxies is not beyond the question. Those galaxies were also excluded from the current consideration.

\section{Results}

\subsection{The properties of galaxies obtained using our measurements}

\begin{figure*}
\begin{center}
\resizebox{1.00\hsize}{!}{\includegraphics[angle=000]{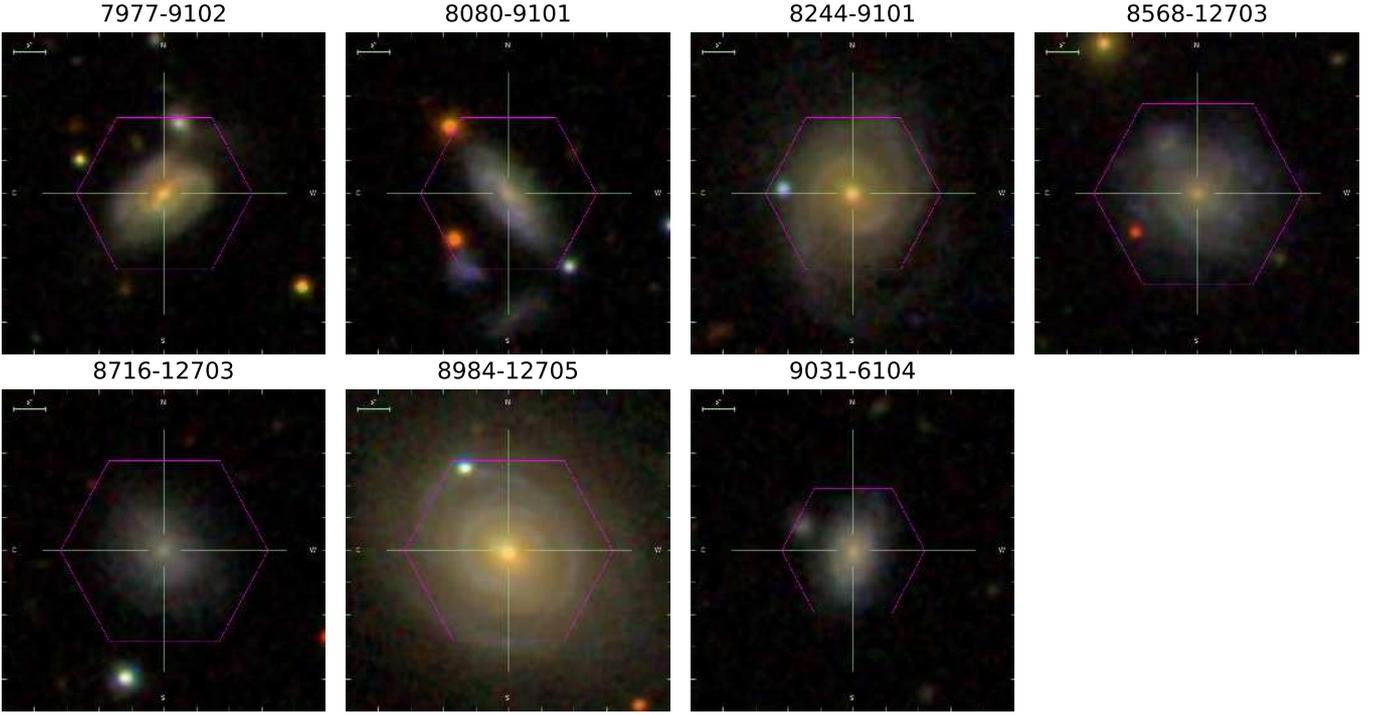}}
\caption{
  Images of galaxies of our sample in the SDSS {\em gri} filters.
}
\label{figure:images}
\end{center}
\end{figure*}

Our final list includes seven galaxies. Fig.~\ref{figure:images} shows the images of the galaxies. 
Table \ref{table:sample} lists the general characteristics of each galaxy. The first column gives its
MaNGA number. The second  column gives other known names. We have indicated the most widely used name for each galaxy, with the following
categories in descending order:  \\
NGC -- New General Catalogue,       \\
UGC -- Uppsala General Catalog of Galaxies,   \\
PGC -- Catalogue of Principal Galaxies.  \\
The right ascension (RA) and declination (DEC) (J2000.0) of each galaxy are given in columns 3 and 4. The right ascension and declination
are taken from the NASA/IPAC Extragalactic Database ({\sc ned}). 
The kinematic inclination angles determined in this paper are listed in column 5.
The spectroscopic stellar masses $M_{sp}$, in solar masses, are reported in column 6. 
The optical radius $R_{25}$ in Kpc determined here are given in column 7. 
The adopted distances (taken from the {\sc ned})  are reported in column 8. 
The galactocentric distances  (normalized to the optical radius $R_{25}$) of the spot position are listed in column 9. 
The BPT type of radiation at the center of the galaxy (0 -- SF, 1 -- intermediate, 2 -- AGN) and the BPT type of radiation in the spot
are given in column 10 and 11, respectively.

The maps of the surface brightness, the line-of-sight velocity, the oxygen abundance, the gas velocity dispersion, and the BPT types of the spectra for the MaNGA
galaxies in our sample are shown in Fig.~\ref{figure:m-7977-09102} --  Fig.~\ref{figure:m-9031-06104}.
The location of the spot is labeled with a circle. It should be noted that the circle is not the boundary of the spot but shows the point spread
function (PSF) of the MaNGA measurements which is estimated to have a full width at half maximum of 2.5 arcsec, or 5 pixels \citep{Bundy2015,Belfiore2017}.
The location of the center of the spot is estimated by eye and is somewhat arbitrary. 
The comments to individual galaxies are given below.

{\em M-7977-09102.}
This is a giant galaxy of mass 10$^{10.935}M_{\odot}$ at a distance of 251.4 Mpc. The M-7977-09102 galaxy has
a close satellite (SDSS J221119.18+121116.0) of mass 10$^{9.489}M_{\odot}$. The separation between  M-7977-09102
and its satellite galaxy on the sky plane is $\sim$14 Kpc (that is comparable to the optical radius of  M-7977-09102, $R_{25}$ = 12.8 Kpc)
and the separation in the line-of-sight velocity is d$V_{los}$ $\sim$ 230 km/s (that is comparable with the variation of the V$_{los}$ across the
image of  M-7977-09102, see panel b of Fig.~\ref{figure:m-7977-09102}). 
The field of view of the MaNGA measurement covers partly the satellite. The enhanced gas velocity dispersion takes place in the region at the edge of
M-7977-09102, which is close to the satellite. The emission lines in the spectra of the spaxels within the spot of the enhanced gas
velocity dispersion are single-peaked, see panels f2 and f3 of Fig.~\ref{figure:m-7977-09102}. This evidences against a false enhancement 
of the lines width due to two separate lines (M-7977-09102 and satellite). 
Some spaxels in the spot show the spectra of  intermediate BPT type.

{\em M-8080-09101.}
This galaxy has a mass of 10$^{9.891}M_{\odot}$, and it is at a distance of 180.3 Mpc. The M-8080-09101 galaxy has a close satellite (SDSS J030922.02-002440.5)
of mass  10$^{9.489}M_{\odot}$. The separation between the  M-8080-09101 galaxy and its satellite in the sky plane is comparable to the optical radius of
 M-8080-09101, and the separation in the line-of-sight velocity is comparable to the variation of the V$_{los}$ across the image of M-8080-09101,
see panel b of Fig.~\ref{figure:m-8080-09101}. The field of view of the MaNGA measurement covers partly the satellite. The enhanced gas velocity dispersion can
be seen in both, the region of  M-8080-09101 which is close to the satellite, and  the region of the satellite which is close to  M-8080-09101.
The lines in the spectra of the spaxels with enhanced gas velocity dispersion are single-peaked, see panels f2 and f3 of Fig.~\ref{figure:m-8080-09101}.
The gas-phase oxygen abundance in the satellite is lower by a factor of around three as compared to the oxygen abundance in  M-8080-09101
(panel c of Fig.~\ref{figure:m-8080-09101}). However, the oxygen abundance at the edge of the satellite close to  M-8080-09101 is higher than
in the rest of the satellite. This can be an indicator of the gas exchange between  M-8080-09101 and its satellite. Unfortunately, the spectra of the
spaxels in the region between M-8080-09101 and the satellite are noisy, which prevents the determination of the oxygen abundances and the further
investigation of a detailed picture of gas exchange between those galaxies.  

{\em M-8244-09101.}
This galaxy has  a mass of 10$^{11.201}M_{\odot}$, and it is at a distance of 241.6.7 Mpc. There is a spot of enhanced gas velocity dispersion
at a galactocentric distance of around 15 Kpc ($\sim$0.8 of the optical radius $R_{25}$, panels d1 and d2 of Fig.~\ref{figure:m-8244-09101}). 
The gas-phase oxygen abundance in the spot is lower in comparison to the abundance at this radius (panels d1 and d2 of Fig.~\ref{figure:m-8244-09101}). 
The lines in the spectra of the spaxels with enhanced gas velocity dispersion are single-peaked, see panels f2 and f3 of Fig.~\ref{figure:m-8244-09101}.

{\em M-8568-12703.}
This galaxy has  a mass of 10$^{10.469}M_{\odot}$ and it is at a distance of 221.2 Mpc. There is a large spot of enhanced gas velocity dispersion at a galactocentric
distance of around 11 Kpc ($\sim$0.8 of the optical radius $R_{25}$, panels d1 and d2 of Fig.~\ref{figure:m-8568-12703}). The spot is also revealed on the surface
brightness and the line-of-sight velocity maps. The lines in the spectra of the spaxels with enhanced gas velocity dispersion are single-peaked,
see panels f2 and f3 of Fig.~\ref{figure:m-8568-12703}.

{\em M-8716-12703.}
This galaxy has  a mass of 10$^{9.770}M_{\odot}$ and it is at a distance of 174.5 Mpc. There is a spot of enhanced gas velocity dispersion at a galactocentric distance
of around 6 Kpc ($\sim$0.7 of the optical radius $R_{25}$, panels d1 and d2 of Fig.~\ref{figure:m-8716-12703}).
The configuration of the radiation distribution in the spot resembles the  configuration of the radiation distribution in a circumnuclear
AGN, that is the innermost region of the AGN-like radiation is surrounded by a ring of  radiation of the intermediate type (panel e of Fig.~\ref{figure:m-8716-12703}).
The spot is not exhibited appreciable on the surface brightness and the line-of-sight velocity maps.  
The lines in the spectra of the spaxels with enhanced gas velocity dispersion are single-peaked, see panels f2 and f3 of Fig.~\ref{figure:m-8716-12703}.

{\em M-8984-12705.}
Also known as NGC~5251, this galaxy is an isolated S0-a (HyperLeda database) galaxy of mass 10$^{11.292}M_{\odot}$ at a distance of 158.3 Mpc.
There is a spot of enhanced gas velocity dispersion at a galactocentric distance of around 12 Kpc ($\sim$0.6 of the optical radius $R_{25}$, panels d1 and d2 of
Fig.~\ref{figure:m-8984-12705}). The gas-phase oxygen abundance in the spot is reduced in comparison to the abundance at this radius (panel c
of Fig.~\ref{figure:m-8984-12705}). The spot is also a prominent feature on the surface brightness map.
The complexity of the H$\alpha$ line profile  provides evidence that the radiation is originated in several clumps, that is several gas fragments from the
satellite are mixed to the interstellar medium of the disk and each fragment produces different enhancements of gas velocity dispersion and/or 
different changes in the line-of-sight velocity. However, the disagreement between H$\alpha$ and H$\beta$ line profiles (see panels f2 and f3
of Fig.~\ref{figure:m-8984-12705}) prevents a solid interpretation of the origin of the interaction. The spaxel spectra in the spot are noisy,
more accurate (with higher spectral resolution) measurements are necessary to make a solid conclusion.  

{\em M-9031-06104.}
This galaxy has a mass 10$^{10.019}M_{\odot}$ and it is at a distance of 185.1 Mpc. The M-9031-06104 galaxy has a close satellite.
The separation between  M-9031-06104 and its satellite exceeds slightly the optical radius of  M-9031-06104 ($R_{25}$ = 8.5 Kpc), 
the line-of-sight velocities of  M-9031-06104 and its satellite are close to each other, Fig.~\ref{figure:m-9031-06104}.
The field of view of the MaNGA measurement covers partly the satellite. The enhanced gas velocity dispersion takes place at the edge of
M-9031-06104 which looks to the satellite. The emission lines in the spectra of the spaxels with enhanced gas velocity dispersion 
are single-peaked, see panels f2 and f3 of Fig.~\ref{figure:m-9031-06104}.

\subsection{The properties of galaxies obtained using the MaNGA Data Analysis Pipeline (DAP) measurements}

\begin{figure}
\begin{center}
\resizebox{1.00\hsize}{!}{\includegraphics[angle=000]{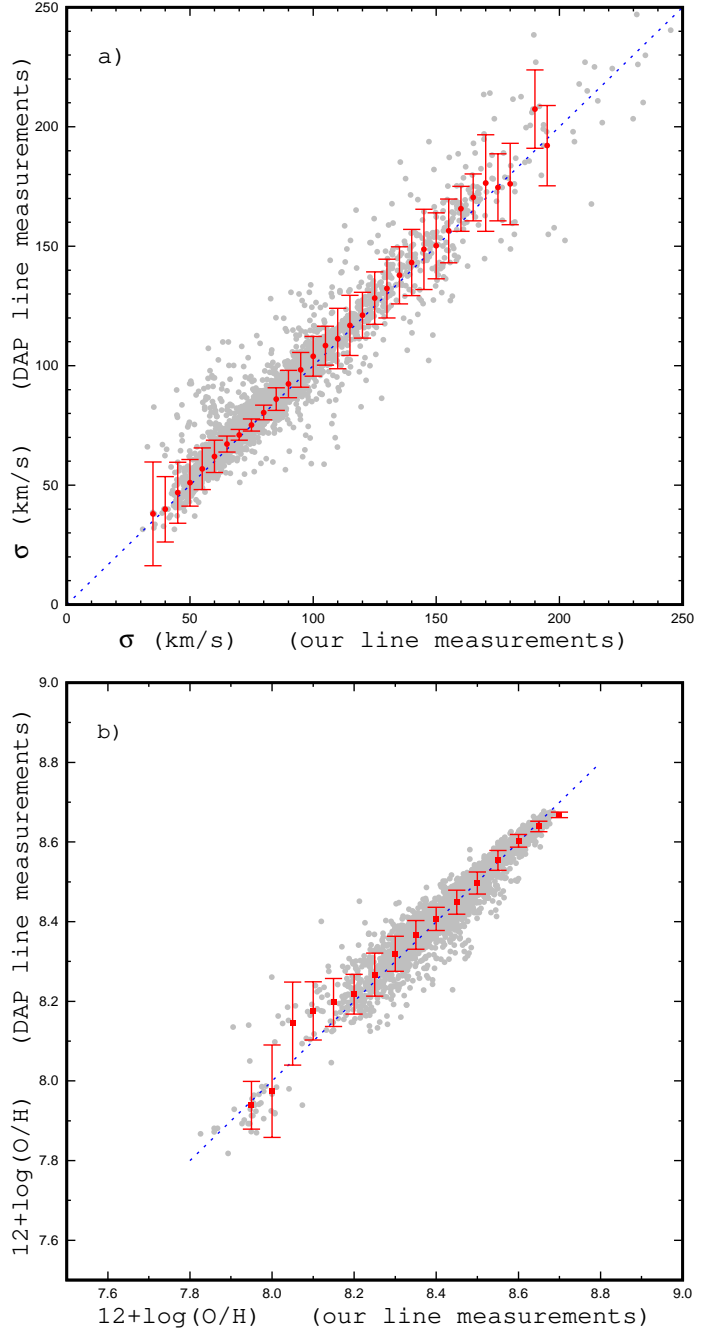}}
\caption{
Comparison between the MANGA-DAP and our measurements for the values of the gas velocity dispersion $\sigma$ and the oxygen abundance O/H.
{\em  Panel a}: the grey points show the gas velocity dispersion of individual spaxels. The red points show the mean values of velocity dispersion in bins of 5 km/s.
{\em  Panel b}: the grey points show the oxygen abundance of individual spaxels.
The red points show the mean values of oxygen abundance in bins of 0.05 dex. In both panels the dotted-line is the one-to-one relation.
}
\label{figure:igordap}
\end{center}
\end{figure}

The emission line parameters for our sample of galaxies are available from the MaNGA data analysis pipeline (DAP) measurements.
We use the datacubes from the most recent publicly available data release DR16, which is based on the DAP version v2.4.3.
The important difference between the DAP measurements and ours is that DAP is fitting all emission lines simultaneously with tied line centers and 
line widths. For each spaxel spectrum, we take the fluxes of the [O\,{\sc ii}]$\lambda\lambda$3727,3729, H$\beta$,
[O\,{\sc iii}]$\lambda$5007, H$\alpha$, and [N\,{\sc ii}]$\lambda$6584 emission lines, the line-of-sight velocity and the gas velocity dispersion
derived from the H$\alpha$ line from the datacube with HYB10 binning scheme. In this case a Voronoi binning of S/N=10 is applied to the stellar spectra which is then
used for the stellar kinematics. The bins are then deconstructed such that the emission-lines fitting is measured on the individual spaxels.

We also take the photometric inclination angle and the position angle of the major photometric axis for each galaxy from the NASA-Sloan Atlas (NSA)
catalog\footnote{http://nsatlas.org}. Both angles are obtained from the Sersic fit to the surface brightness profile in the r band.

The maps of the line-of-sight velocity, the oxygen abundance, the gas velocity dispersion, and the BPT types of the spectra for our sample of  MaNGA
galaxies were constructed using the DAP measurements. The maps are shown in Fig.~\ref{figure:m-7977-09102-dap} --  Fig.~\ref{figure:m-9031-06104-dap}.
A visual comparison between the maps based on our measurements (Fig.~\ref{figure:m-7977-09102} --  Fig.~\ref{figure:m-9031-06104}) and the maps based on the DAP measurements
(Fig.~\ref{figure:m-7977-09102-dap} --  Fig.~\ref{figure:m-9031-06104-dap}) shows that there is a satisfactory agreement.
It should be emphasized however that the DAP measurements within the spots of two galaxies (M-8244-09101 and M-8984-12705) are not available.

Here we show a quantitative comparison between  spaxel properties  based on ours and DAP measurements.   
Panel a of Fig.~\ref{figure:igordap} shows the comparison between the gas velocity dispersion $\sigma$ from DAP vs. our measurements. 
The grey points correspond to 13219 individual spaxels in seven galaxies, while the red points are the mean values of velocity dispersion in bins of 5 km/s.
The bars denote the mean value of the scatter of the $\sigma$ in each bin. Only bins containing five or more points are shown.
The mean value of the scatter of the gas velocity dispersion around the one-to-one relation is $\sim$5 km/s.
This value can be used to specify the uncertainty in the values of the gas velocity dispersion.
Inspection of panel a of Fig.~\ref{figure:igordap} shows that our values of the gas velocity dispersion and the values from the DAP measurements are in satisfactory agreement.
There is no appreciable systematic difference between $\sigma_{our}$ and $\sigma_{DAP}$ values over the whole interval of $\sigma$, that is, the $\sigma_{our}$ -- $\sigma_{DAP}$
diagram follows well the one-to-one relation. 

Panel b of Fig.~\ref{figure:igordap} shows the comparison between the determined oxygen abundances from the DAP and our measurements.
The grey points are the data of 6979 individual spaxels in seven galaxies. The red points indicate the mean values of oxygen abundances in bins of 0.05 dex. 
The mean value of the scatter of the oxygen abundance around equal-value line is $\sim$0.025 dex.   
This value can serve as an estimation of the uncertainty in the values of the oxygen abundance.
Inspection of panel b of Fig.~\ref{figure:igordap} shows that values of the oxygen abundances determined from our and the DAP measurements are in satisfactory
agreement over the whole interval of metallicities.

\begin{figure*}
\begin{center}
\resizebox{1.00\hsize}{!}{\includegraphics[angle=000]{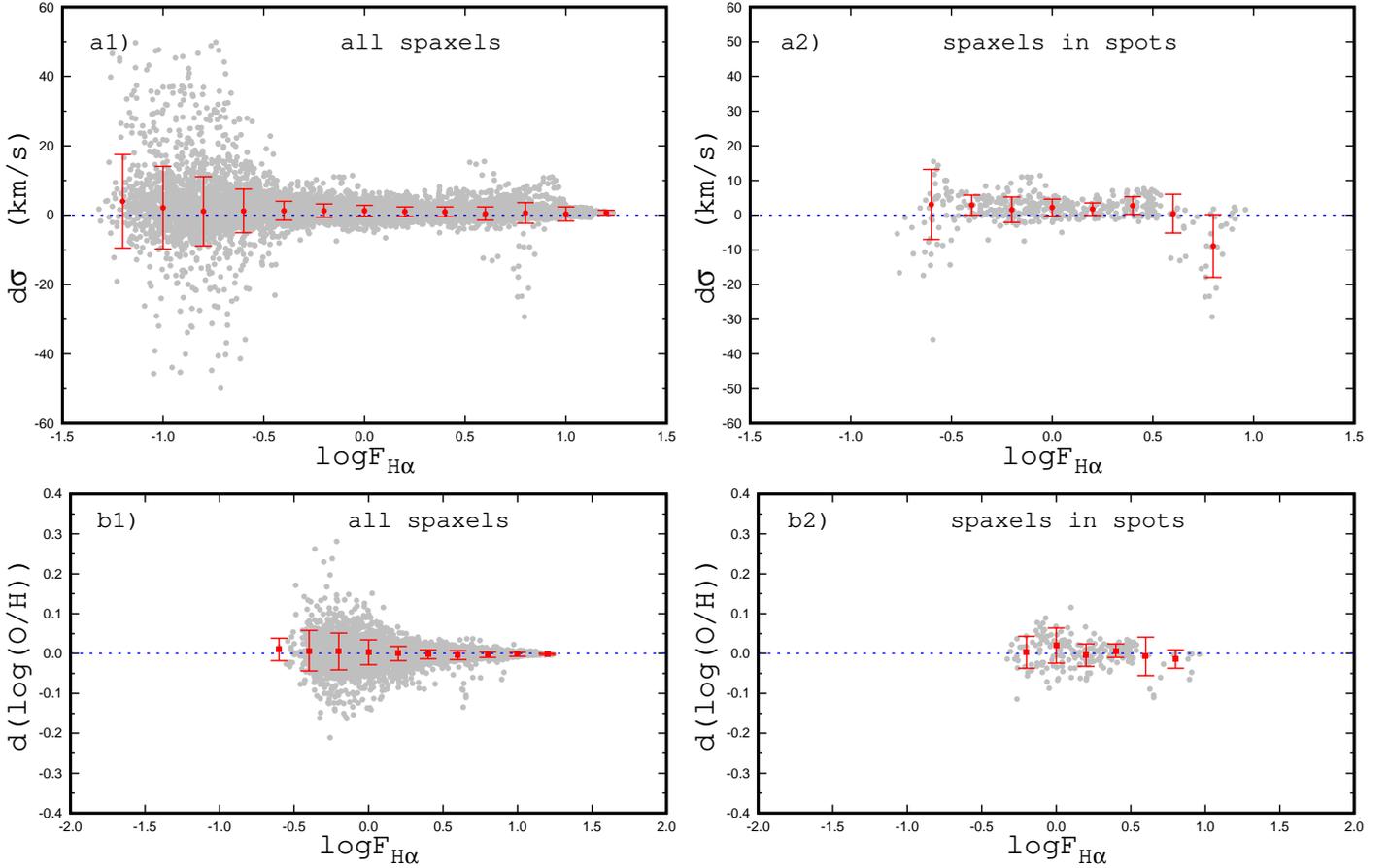}}
\caption{
Difference between the DAP and our measurements of the gas velocity dispersion $\sigma$ and the oxygen abundance O/H.
{\em  Panel a1}: the grey points show the difference d$\sigma$ = $\sigma_{DAP}$ -- $\sigma_{our}$ as a function of H$_{\alpha}$ flux for all the individual spaxels
in seven galaxies. The red points show the mean values of the difference in bins of 0.2 dex in H$_{\alpha}$ flux. 
{\em  Panel a2} shows the same as panel a1 but for the spaxels in the spots. 
{\em  Panel b1}: the grey points show the difference d(O/H) = log(O/H)$_{DAP}$ -- log(O/H)$_{our}$  as a function of H$_{\alpha}$ flux for all the individual spaxels.
The red points show the mean values of difference in bins of 0.2 dex in H$_{\alpha}$ flux.
{\em  Panel b2} shows the same as panel b1 but for the spaxels in the spots. 
}
\label{figure:lhadsvdoh}
\end{center}
\end{figure*}

Panel a1 of Fig.~\ref{figure:lhadsvdoh} shows the difference between the values of the gas velocity dispersion from  the DAP and our measurements  
d$\sigma$ = $\sigma_{DAP}$ -- $\sigma_{our}$ as a function of H$_{\alpha}$ flux for all the spaxels. The measured (non-deredddened) H$\alpha$ fluxes in units of
10$^{-17}$ erg/s/cm$^{2}$/spaxel is presented. The grey points show the d$\sigma$ for the individual spaxels, while the red points are the mean values of d$\sigma$
in bins of 0.2 dex in H$_{\alpha}$ flux. The bars denote the mean value of the scatter of the d$\sigma$ in each bin.
Inspection of panel a1 of  Fig.~\ref{figure:lhadsvdoh} shows that the mean values of d$\sigma$ are close to zero over the whole range  of H$\alpha$ fluxes, while 
the scatter in d$\sigma$ is larger for low H$\alpha$ fluxes.

It is interesting to examine the values of the d$\sigma$ for the spaxels in the spots of enhanced gas velocity dispersion. It is impossible to determine the exact
location of the center of the spot and its border since the measurements are usually available for only the part of the spot (see, for example,  Fig.~\ref{figure:m-8080-09101}).
Then we consider the regions outlined by rings in  Fig.~\ref{figure:m-7977-09102} -- Fig.~\ref{figure:m-9031-06104} as the spots and 
examine the values of the d$\sigma$ for the spaxels within those regions.
Panel a2 of Fig.~\ref{figure:lhadsvdoh} shows the d$\sigma$ as a function of H$_{\alpha}$ flux for the spaxels in the spots. The grey points denote the d$\sigma$ for
the individual spaxels, while the red points are the binned mean values of d$\sigma$.  The bars denote the scatter of the d$\sigma$ in each bin.
Inspection of panel a2 of  Fig.~\ref{figure:lhadsvdoh} shows that the values of the d$\sigma$ in the spaxels within the spots are similar to that in the spaxels
outside the spots. The values of the d$\sigma$ for the majority of the spaxels in the spots are less than $\sim$6 km/s for the galaxies considered.

Panel b1 of Fig.~\ref{figure:lhadsvdoh} shows the difference between the values of the oxygen abundance determined from the DAP and our measurements  
d(O/H) = log(O/H)$_{DAP}$ -- log(O/H)$_{our}$ as a function of the H$_{\alpha}$ flux for all the spaxels. The grey points show the d(O/H) for the individual spaxels, while the
red points are the mean values of d(O/H) in bins of 0.2 dex in  H$_{\alpha}$ flux. Inspection of panel b1 of  Fig.~\ref{figure:lhadsvdoh} shows again that mean values of
the d(O/H)  are close to zero over the whole range  of H$\alpha$ fluxes, while the scatter in d(O/H) is larger at low H$\alpha$ fluxes.
Panel b2 of Fig.~\ref{figure:lhadsvdoh} shows the d(O/H) as a function of H$_{\alpha}$ flux for the spaxels in the spots. 
Inspection of this panel shows that the values of the d(O/H) of the spaxels within the spots are similar to those of the spaxels
outside the spots. The values of the d(O/H) for the majority of the spaxels in the spots are less than $\sim$0.05 dex for the galaxies considered.

\begin{figure}
\begin{center}
\resizebox{1.00\hsize}{!}{\includegraphics[angle=000]{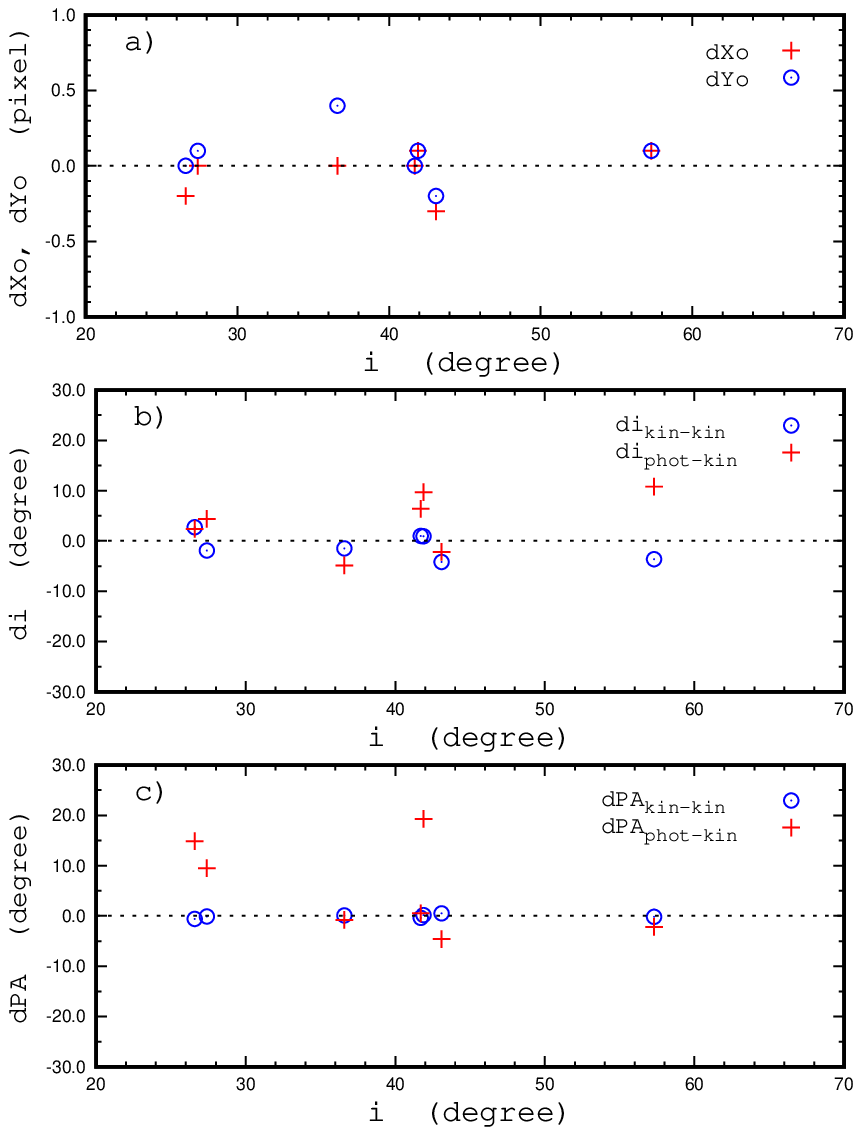}}
\caption{
Comparison between the different sets of geometrical parameters for our sample of galaxies.
{\em  Panel a}: differences between the coordinates of the kinematical center of galaxies dXo and dYo  derived using DAP and our measurements. 
{\em  Panel b}: differences between two determinations of the kinematical inclination angle, and between the kinematic and photometric inclination angles.
{\em  Panel c}: differences between two determinations of the position angle of the major kinematical axis, and between the position angles of the major
axis and the major photometric axis. See text for details.
}
\label{figure:dangles}
\end{center}
\end{figure}

The geometrical parameters of galaxies (coordinates of the rotating center Xo and Yo, the position angle of the major kinematic angle PA$_{kin}$,
and the inclination angle $i_{kin}$) were derived using both, our and DAP line-of-sight velocity fields. In addition, the photometric inclination angle $i_{phot}$
and the position angle of the major photometric axis PA$_{phot}$ for each galaxy were taken from the DAP data. Then we can compare the two sets of  coordinates for the rotating center (e.g. dXo = Xo$_{DAP}$ - Xo$_{our}$) and three sets of inclination and position angles (e.g. dPA$_{kin-kin}$ =
PA$_{kin,DAP}$ - PA$_{kin,our}$ and dPA$_{phot-kin}$ = PA$_{phot}$ - PA$_{kin,our}$).
Panel a of Fig.~\ref{figure:dangles} shows the dXo and dYo as a function of the kinematic inclination angle.
Inspection if this panel shows that the values of dXo and dYo are usually within $\sim$0.2 and $\sim$0.4 spaxel for the galaxies
considered here.
Panel b of Fig.~\ref{figure:dangles} shows the values of the d$i_{kin-kin}$ and d$i_{phot-kin}$  as a function of the kinematic inclination angle.
Inspection of this panel shows that the values of d$i_{kin-kin}$ are within $\sim$4$\degr$  while the values of d$i_{phot-kin}$
can reach  up to $\sim$12$\degr$.
Panel c of Fig.~\ref{figure:dangles} shows the values of the dPA$_{kin-kin}$ and dPA$_{phot-kin}$  as a function of the kinematic inclination angle.
Inspection of this panel shows that the values of dPA$_{kin-kin}$ are within $\sim$1$\degr$  while the values of dPA$_{phot-kin}$
can reach  up to $\sim$20$\degr$.
The geometrical parameters of galaxies derived from the analysis of the line-of-sight velocity field are rather robust. 
This can be attributed to the fact that an iterative procedure is used in the determination of the geometrical parameters and rotation curve,
that is, the points with large deviations from the rotation curve are rejected.  

Thus, the properties of the gas in the spaxels derived using our and the DAP measurements are in satisfactory agreement. There is no systematic difference between the 
$\sigma_{our}$ and $\sigma_{DAP}$ values over the whole interval of $\sigma$, that is the $\sigma_{our}$ -- $\sigma_{DAP}$ diagram follows well the one-to-one relation. 
There is not a significant difference between (O/H)$_{our}$ and (O/H)$_{DAP}$ values over the whole interval of metallicity, that is the
(O/H)$_{our}$ -- (O/H)$_{DAP}$ diagram  also follows the one-to-one relation. Hence, the  quantities based on our measurements are used below.

\section{Discussion}

The following questions are discussed in this Section. First we discuss the evidences in favor of that the spots of enhanced gas velocity dispersion
in six galaxies of our sample can be caused by  (minor) mergers or interactions. 
Then we examine whether the presence of the spot of enhanced $\sigma$ is accompanied by the enhancement of the median value of the gas velocity dispersion
in the host galaxy, that is whether the minor interaction (or merger) results in the disturbance of the gas velocity dispersion in a localize region or in the galaxy as whole. 
Finally, we discuss the spot of enhanced $\sigma$ in the galaxy M-8716-12703 with AGN-like  configuration of the radiation distribution. This spot is compared
with the AGN in the galaxy M-8984-12705.

\subsection{Minor merger or interaction as a reason for the spot of enhanced $\sigma$}

Three galaxies in our sample (M-7977-09102, M-8080-09101, and M-9031-06104) have a very close satellite
(see the images for those galaxies in Fig.~\ref{figure:images}, and the maps of the surface brightness and line-of-sight velocity distributions across the images
in Fig.~\ref{figure:m-7977-09102}, Fig.~\ref{figure:m-8080-09101}, and Fig.~\ref{figure:m-9031-06104}, as well as the comments for those galaxies above).
The separation in the sky plane is comparable
to the optical radius of the galaxy in each case, and the separation in the line-of-sight velocity is comparable to the variation of the V$_{los}$ across the image
of the galaxy.  The spots of the enhanced $\sigma$ in those galaxies are located at the edge of the galaxy close to the satellite. This suggests that
the spots of the enhanced $\sigma$ in those galaxies are caused by the interaction with the satellite (gas flow from the satellite).

The spots of  enhanced $\sigma$ in four other galaxies (M-8244-09101, M-8568-12703, M-8716-12703, and M-8984-12705) 
are located within the optical radius, at the fractional radii from $\sim$0.6 to $\sim$0.8  optical
radius. The spots of the enhanced $\sigma$ in three of those galaxies are related to the bright spots in the photometric $B$ band 
  (see images for those galaxies in Fig.~\ref{figure:images}, and the maps of the surface brightness across the images for those galaxies in
  Fig.~\ref{figure:m-8244-09101}, Fig.~\ref{figure:m-8568-12703}, and Fig.~\ref{figure:m-8984-12705}, as well as the comments for  those galaxies above).
This could be an indicator
of a projection of a satellite on the galaxy. It was noted above that the projection of a satellite on the galaxy takes indeed place in three galaxies
(M-8553-09102, M-9029-12705, and M-9049-12701) from our preliminary list. Two sets of the emission lines are present in the spectra of the spaxels within the spots
in two galaxies M-8553-09102 and M-9029-12705, that is, both set of emission lines of the galaxy and its satellite are detected. The emission lines in the spectra
of the spaxels within a spot in the galaxy M-9049-12701 are double-peaked, see panels f2 and f3 of Fig.~\ref{figure:m-9049-12701}. Those galaxies were
excluded from the current consideration. Regarding the three galaxies in our final list with enhanced $\sigma$ spots associated to the bright $B$ band spots
we find the following. A minor (if any) separation  in the line-of-sight velocity between the spot and the surroundings, 
as well as the single-peaked emission lines in the spaxel spectra in the spot suggest that the emission lines in the spectra of the spaxels in the enhanced
$\sigma$ spot originate in a single region located within (or at least very close to) the disk. Thus either the satellite is already captured
and is in the disk of the galaxy, or  the region of  enhanced $\sigma$ is caused by the interaction with a satellite (gas infall from the
satellite onto the disk of the galaxy). The oxygen abundances in the spots of enhanced $\sigma$ in these three galaxies are reduced,
this is a strong evidence in favor of that the low-metallicity gas from the satellite is mixed with the gas of the galactic disk.

The spectra of the spaxels within a spot are usually H\,{\sc ii}-region-like. This suggests that the gas infall onto the galaxy
does not necessarily result in appreciable shocks.  In contrast, the spot of enhanced $\sigma$ in the galaxy M-8716-12703 is
associated with an off-centered AGN-like  configuration of the radiation distribution,  in the sense that the innermost region of
the AGN-like radiation is surrounded by a ring of radiation of the intermediate type. There are no evident signs of  gas infall
onto this galaxy. One can suggest that the spot of enhanced $\sigma$ in the galaxy M-8716-12703 is different in  origin or
that the type the interaction (gas infall onto the galaxy) in this case differs from that in other galaxies. 

It is known that the gas velocity dispersion in the giant  H\,{\sc ii} region correlates with the H$\alpha$ luminosity or diameter, in the sense that the $\sigma$ is higher
in the H\,{\sc ii} region with a higher H$\alpha$ luminosity or larger diameter \citep{Terlevich1981,FernandezArenas2018}. There is the minimum value of the gas velocity
dispersion in H\,{\sc ii} regions of a given H$\alpha$ luminosity or diameter for  H\,{\sc ii} regions with a H$\alpha$ luminosity higher than $\sim$10$^{38}$ erg/s
\citep{Relano2005,ZaragozaCardiel2015,AmbrocioCruz2016}. The integral emission line profile of a giant  H\,{\sc ii} region measued at low resolution is rather smooth
and well fitted by a Gaussian, while the observations at high spatial and spectral resolution reveal that a giant  H\,{\sc ii} region breaks up into many discrete components
\citep{Chu1994,Bresolin2020}.  The H\,{\sc ii} regions from interacting galaxies show, on average, higher gas velocity dispersions than those from isolated galaxies
\citep{ZaragozaCardiel2015}. The diameters of the largest giant H\,{\sc ii} regions in galaxies are $\sim$1 kpc \citep{Kennicutt1988,ZaragozaCardiel2015}.
The minimum observed size of the spots of enhanced $\sigma$ considered is $\sim$6 pixels (only part of the spots is usually measured). The distances of galaxies of our
sample are from 158 to 251 Mpc (see Table 1), which give a scale of $\sim$0.4-0.6 kpc/pixel, meaning that the observed sizes of the spots are larger than $\sim$2.5 kpc. 
Thus, the sizes of the spots of enhanced $\sigma$ we considered exceed the sizes of the largest known  H\,{\sc ii} regions. However, the observed sizes of the spots are
comparable to the point spread function (PSF) of the MaNGA measurements which is estimated to have a full width at half maximum of 5 pixels \citep{Bundy2015,Belfiore2017}.
Therefore, the measurements of the spots with a higher resolution and higher S/N are necessary to make accurate estimations of their sizes (and other characteristics)
and to perform the quantitative comparison between spots and giant H\,{\sc ii} regions.
  
\begin{figure}
\begin{center}
\resizebox{1.000\hsize}{!}{\includegraphics[angle=000]{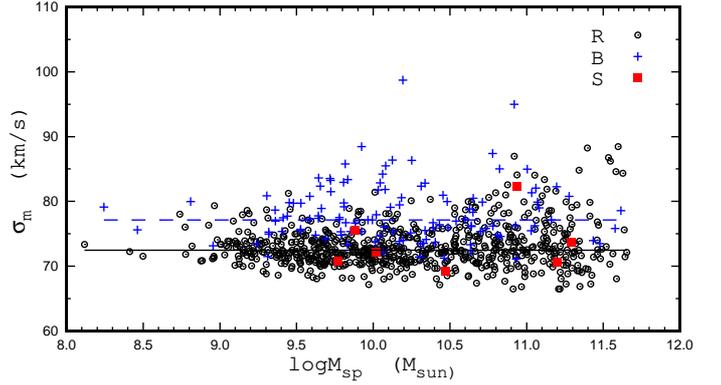}}
\caption{
Median values of gas velocity dispersion $\sigma_{m}$ in galaxies. 
The black circles show $\sigma_{m}$ as a function of the spectroscopic stellar mass $M_{sp}$ for
individual galaxies with the R distribution, the blue plus signs denote galaxies with the B distribution,
and red squares  galaxies with the spots of enhanced $\sigma$.
The solid line shows the median value of $\sigma_{m}$ in galaxies with the R distribution, and
the dashed line shows the median value of $\sigma_{m}$ for galaxies with the B distribution.
The $\sigma_{m}$ for galaxies with R and B distributions are from \citet{Pilyugin2021}.
}
\label{figure:msp-svm}
\end{center}
\end{figure}

\subsection{Relation between presence of spot and enhancement of $\sigma$ in whole galaxy?}

Above we have discussed evidences in favor  that the off-centered spots of  enhanced $\sigma$ in galaxies can be caused by (minor) mergers or interactions. 
Here we consider the median values of the gas velocity dispersion $\sigma_{m}$ in galaxies as a whole aiming to examine whether there is a
 difference between galaxies with and without spots of  enhanced $\sigma$. The red squares in Fig.~\ref{figure:msp-svm}  show the median
values of the observed gas velocity dispersion $\sigma_{m}$ as a function of the spectroscopic stellar mass $M_{sp}$ for our sample of galaxies
with the spots of  enhanced $\sigma$. A sample of galaxies from \citet{Pilyugin2021} is used for comparison. Since the median value of
the gas velocity dispersion $\sigma_{m}$ in a galaxy is related to the type of  distribution of the gas velocity dispersion $\sigma$ across
its image (in the sense that the $\sigma_{m}$ in galaxies with B distribution is higher by around 5 km/s, on average, than that of galaxies
with R distribution), then we distinguish the galaxies with R and B distributions in the comparison sample. Fig.~\ref{figure:msp-svm}
shows the median value of the observed gas velocity dispersion $\sigma_{m}$ as a function of the spectroscopic stellar mass $M_{sp}$ for galaxies
with the R distribution (black circles) and for galaxies with the  B distribution (blue plus signs). The solid line in Fig.~\ref{figure:msp-svm}
denotes the median value  of the $\sigma_{m}$ for the R distribution (72.45 $\pm$ 3.24 km/s), while the dashed
line corresponds to the B distribution (77.15 $\pm$ 4.66 km/s).

Examination of  Fig.~\ref{figure:msp-svm} shows that the $\sigma_{m}$ in six out of seven galaxies with the spots of enhanced $\sigma$ is close
to the typical value of the $\sigma_{m}$ in galaxies without the spots of enhanced $\sigma$. On the other hand, the value of one galaxy with the spot
of enhanced $\sigma$ (M-7977-09102) is slightly higher than the typical value of the $\sigma_{m}$ in galaxies without the
spots of enhanced $\sigma$. This can be considered as an evidence in favor of that the minor interaction (or merger) results usually in the
disturbance of the gas velocity dispersion in a localize region, but not in the galaxy as whole. 

\begin{figure}
\begin{center}
\resizebox{1.0\hsize}{!}{\includegraphics[angle=000]{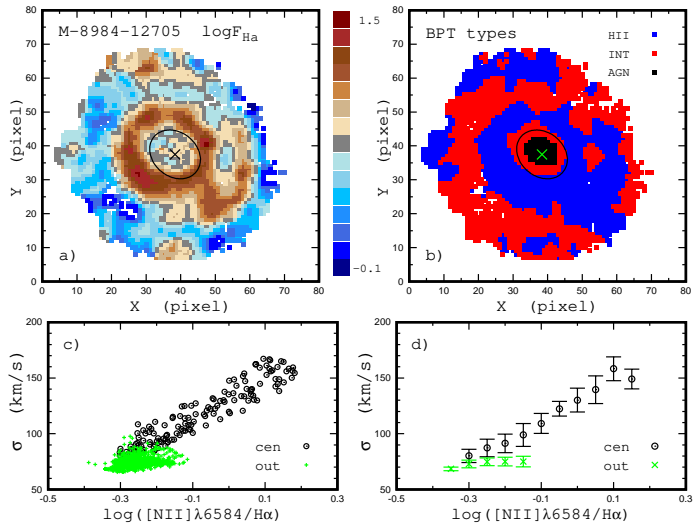}}
\caption{
  Properties of the galaxy M-8984-12705 with central AGN and extranuclear LINERs.
  {\em Panel} $a$: distribution of the H$\alpha$ flux (in logarithmic scale) in the spaxel spectra across the image of the galaxy.
  The value of the flux is color-coded. The cross shows the kinematic center of the galaxy, the ellipse outlines the adopted circumnuclear region.
  {\em Panel} $b$:  locations of the spaxels with spectra of different BPT types (the AGN-like,  H\,{\sc ii}-region-like,
  and intermediate) on the image of the galaxy.
  {\em Panel} $c$: relation between gas velocity dispersion $\sigma$ and [N\,{\sc ii}]$\lambda$6584/H$\alpha$ line ratio for individual spaxels with intermediate
  and AGN-like  spectra in the circumnuclear (dark circles) and in the extranuclear (green plus signs) regions of the galaxy M-8984-12705.
  {\em Panel} $d$: median values of $\sigma$ for  spaxels in bins of 0.05 dex in [N\,{\sc ii}]$\lambda$6584/H$\alpha$ for spaxels in the circumnuclear
  (dark circles) and in the extranuclear (green crosses) regions.
}
\label{figure:xbpt-sv}
\end{center}
\end{figure}

\subsection{Spot in the galaxy M-8716-12703: comparison to AGN in galaxy M-8984-12705}

The galaxy M-8984-12705 with off-centered spot of enhanced $\sigma$ from our sample shows central AGN and an extended extranuclear LINERs
(Low Ionization Nuclear Emission line Regions). The different sources of the ionizing photons in the LINERs are discussed.
It was suggested \citep[e.g.,][]{Stasinska2006,Stasinska2008,Sarzi2010,YanBlanton2012,Singh2013}
that the radiation of the hot, low-mass evolved (post-asymptotic giant branch) stars (HOLMES) can be the source of the ionizing photons in the LINERs.
It was argued \citep{Heckman1980,DopitaSutherland1995,Ho2014,Molina2018} that a LINER can be also excited by the shocks caused by different phenomena (jets, galactic winds,
galaxy-galaxy interactions). 
\citet{Monreal2006,Monreal2010} have studied an extended extranuclear LINERs in ultraluminous infrared galaxies. They revealed a positive correlation between the gas velocity
dispersion and the diagnostic emission line ratios of the BPT diagram.  This correlation is interpreted as a signature of shock excitation,
where shocks are driven by galaxy mergers or interactions \citep{Monreal2006,Monreal2010,Rich2014,Rich2015}.

Panel a in Fig.~\ref{figure:xbpt-sv} shows the distribution of the flux in the  H$\alpha$ line across the image of the galaxy M-8984-12705 (NGC~5251).
We use the de-reddened flux in the  H$\alpha$ line per spaxel $F_{{\rm H}\alpha}$ in units 10$^{-17}$ erg/s/cm$^2$/spaxel (not corrected for the galaxy inclination)
since only the relative variation of the surface brightness in the H$\alpha$ line within the galaxy is considered. Further, since the current star formation rate
is commonly estimated from the H$\alpha$ luminosity of a galaxy  using the calibration relation of \citet{Kennicutt1998} then the panel a in Fig.~\ref{figure:xbpt-sv}
can be interpreted as a map of the relative variation of the current star formation rate (in arbitrary units) within the galaxy.  
A prominent feature in the distribution of the H$\alpha$ flux in the image of the galaxy M-8984-12705 is the presence of a bright ring.

Panel b in Fig.~\ref{figure:xbpt-sv} shows the locations of the spaxels with spectra of different BPT types (the AGN-like,  H\,{\sc ii}-region-like,
and intermediate) on the image of the galaxy M-8984-12705. 
A comparison between panel a and panel b in Fig.~\ref{figure:xbpt-sv} shows that the spaxel spectra are  H\,{\sc ii}-region-like in the areas of
high star formation rate, and that the spectra are intermediate or  AGN-like types correspond to areas of
low  H$\alpha$ flux.  The galaxy M-8984-12705 hosts the AGN at its center and extranuclear LINERs-type regions. 

Next, we examine the relationships between the gas velocity dispersion $\sigma$ and [N\,{\sc ii}]$\lambda$6584/H$\alpha$ line ratio for the central AGN and the extranuclear
LINERs  in the M-8984-12705 galaxy.
Panel d of Fig.~\ref{figure:m-8984-12705} shows that the observed  gas velocity dispersion $\sigma$ decreases
with galactocentric distance up to some radius within this galaxy and remains approximately constant beyond this radius. The radius where the break occurs is adopted as the
radius of the central AGN. The ellipses in panel a and b of Fig.~\ref{figure:xbpt-sv} outline the central AGN in the image of the galaxy M-8984-12705.

Panel c in Fig.~\ref{figure:xbpt-sv} shows the relation between gas velocity dispersion $\sigma$ and [N\,{\sc ii}]$\lambda$6584/H$\alpha$ line ratio for individual
spaxels in the central AGN (dark circles) and in the extranuclear (green plus signs) LINERs of M-8984-12705.
Panel d shows the median values of $\sigma$ for those spaxels in bins of 0.05 dex in [N\,{\sc ii}]$\lambda$6584/H$\alpha$ for spaxels in the central AGN 
(dark circles) and in the extranuclear (green crosses) LINERs.
Examination of Fig.~\ref{figure:xbpt-sv} shows a correlation between the gas velocity dispersion $\sigma$ and the [N\,{\sc ii}]$\lambda$6584/H$\alpha$ line ratio that
takes place only for the central AGN in the  M-8984-12705 galaxy. 
According to  \citet{Monreal2006,Monreal2010} and \citet{Rich2014,Rich2015}, the lack of an appreciable correlation  between gas velocity dispersion $\sigma$ and
[N\,{\sc ii}]$\lambda$6584/H$\alpha$ line ratio for extranuclear LINERs can be considered as evidence that  shocks do not play an appreciable role in their excitation. 
This is in line with our above conclusion that a (minor) merger or interaction does not necessary result in appreciable shocks.

\begin{figure}
\begin{center}
\resizebox{1.0\hsize}{!}{\includegraphics[angle=000]{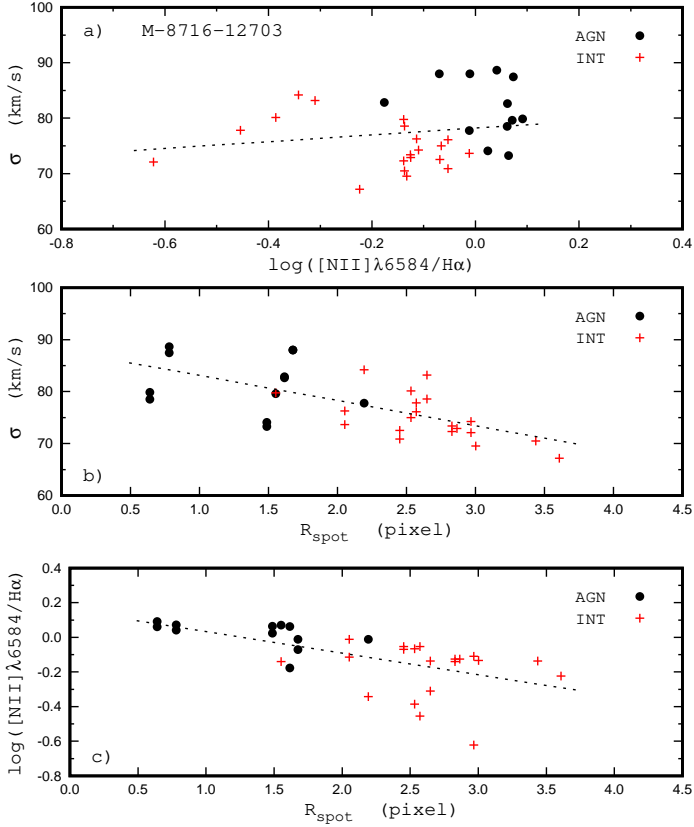}}
\caption{
  Properties of the off-centered AGN-like spot in the galaxy M-8716-12703.
  {\em Panel} $a$: relation between gas velocity dispersion $\sigma$ and [N\,{\sc ii}]$\lambda$6584/H$\alpha$ line ratio for individual spaxels with the 
  AGN-like (circles) and the intermediate (plus signs)  spectra in the spot. 
  {\em Panel} $b$: gas velocity dispersion as a function of the distance to the center of the spot $R_{spot}$ for those spaxels. 
  {\em Panel} $c$:  [N\,{\sc ii}]$\lambda$6584/H$\alpha$ line ratio as a function of the distance to the center of the spot. 
}
\label{figure:m-8716-12703-spot}
\end{center}
\end{figure}

Here we examine properties of the off-centered AGN-like spot in M-8716-12703.
The correlation between the gas velocity dispersion $\sigma$ and the [N\,{\sc ii}]$\lambda$6584/H$\alpha$ line ratio for individual spaxels in the spot for M-8716-12703
is only marginal (if any), the correlation coefficient is equal to 0.176, panel a in Fig.~\ref{figure:m-8716-12703-spot}. 
At the same time, both the gas velocity dispersion $\sigma$ and the [N\,{\sc ii}]$\lambda$6584/H$\alpha$ line ratio show a systematic change with  the distance
to the center of the spot. The correlation coefficient of the relation between the gas velocity dispersion $\sigma$ and the distance
to the center of the spot is equal to --0.66, panel b in Fig.~\ref{figure:m-8716-12703-spot}. 
The correlation coefficient of the relation between the [N\,{\sc ii}]$\lambda$6584/H$\alpha$ line ratio and the distance
to the center of the spot is equal to --0.60, panel c in Fig.~\ref{figure:m-8716-12703-spot}. 
Thus, the properties of the off-centered AGN-like spot in the galaxy M-8716-12703 are controversial. From one hand, the radiation distribution configuration
of the off-centered spot in the galaxy M-8716-12703 is similar to that of the central AGN in the galaxy M-8984-12705, in both cases the innermost region of
the AGN-like radiation is surrounded by a ring of radiation  of the intermediate type. On other hand, the 
relation between the gas velocity dispersion $\sigma$ and the [N\,{\sc ii}]$\lambda$6584/H$\alpha$ line ratio for individual spaxels in the spot in M-8716-12703,
is similar to that of the  extranuclear LINER in the galaxy M-8984-12705, and differs strongly from that for the central AGN in the galaxy M-8984-12705.

The supernova remnants (SNRs) spectra \citep[e.g.,][]{Long2019} are of the AGN-like or the intermediate BPT types. The SNRs exibit the gas velocity dispersion higher
than that in  H\,{\sc ii} regions \citep[e.g.,][]{Points2019}. Those characteristics of SNRs resemble that  of the off-centered AGN-like spot in the galaxy M-8716-12703.  
However, the typical diameter of SNR popolation is 20-40 pc, the SNRs of diameters greater than 100 pc are rare among known SNRs  
\citep{Franchetti2012,Bozzetto2017,Long2020}, while the observed diameter of the off-centered AGN-like spot in M-8716-12703 is $\sim$2.5 kpc (with a scale of $\sim$0.42 kpc/pixel
at the distance 174.5 kpc).  Thus the size of the off-centered AGN-like spot in M-8716-12703 is much larger than that of the SNR.

\section{Conclusions}

We find off-centered spots of enhanced gas velocity dispersion $\sigma$ in a subsample of MaNGA galaxies.
We produce and analyze the distributions of the surface brightness, the line-of-sight gas velocity,
the oxygen abundance, the gas velocity dispersion, and the  Baldwin-Phillips-Terlevich (BPT) types 
of the spaxel spectra in seven MaNGA galaxies aiming to examine the origin of the off-centered spots
of enhanced $\sigma$.

We found that the origin of the spots of enhanced gas velocity dispersion in six galaxies can be
attributed to an interaction (or merger) with a close satellite.
Three galaxies in our sample have a very close satellite (the separation in the sky plane is comparable to the optical radius of the galaxy). The spots of the enhanced
$\sigma$ in those galaxies are located at the edge of the galaxy close to the satellite.
The spots of the enhanced $\sigma$ in three other galaxies are related to the bright spots in the photometric $B$ band within the galaxy. This can be an indicator
of a projected satellite on the line of sight of the galaxy. The oxygen abundances in the spots in these three galaxies are reduced.
 This suggests that the low-metallicity gas from the satellite is mixed with the interstellar medium of the disk,

The spectra of the spaxels within a spot are usually H\,{\sc ii}-region-like. This suggests that the gas
infall onto the galaxy does not necessarily result in appreciable shocks. The  galaxy  M-8984-12705 shows an off-centered spot of enhanced
$\sigma$ and extended extranuclear LINERs regions. The lack of an appreciable correlation  between the
gas velocity dispersion $\sigma$ and the [N\,{\sc ii}]$\lambda$6584/H$\alpha$ line ratio for extranuclear LINERs
evidences that  shocks do not play an important role in their excitation. This also suggests that  (minor)
mergers or interactions does not necessary result in appreciable shocks.

In contrast, the spot of enhanced $\sigma$ in the galaxy M-8716-12703 is associated with an off-centered AGN-like
configuration of the radiation distribution,  in the sense that the innermost region of the AGN-like radiation is
surrounded by a ring of radiation of the intermediate type. There is no evident sign of  gas infall onto this galaxy.
One can suggest that the spot of enhanced $\sigma$ in the galaxy M-8716-12703 has a different origin, or that the
characteristics of the interaction (gas infall onto the galaxy) in this case differs from that in other galaxies. 

\section*{Acknowledgements}

We are grateful to the referee for his/her constructive comments. 
L.S.P acknowledges support within the framework of the program of the NAS of
Ukraine “Support for the development of priority fields of scientific
research” (CPCEL 6541230)".  
BC, JN and MG are supported by AYA2014\,-\,58861\,-\,C3\,-\,1\,-\,P.   
I.A.Z acknowledges support by the National Academy of Sciences of Ukraine
under the Research Laboratory Grant for young scientists No. 0120U100148.  
AMP is supported by AYA2017\,–\,88007\,–\,C3\,–\,2, MDM-2017-0737 
(Unidad de Excelencia María de Maeztu, CAB).   
JMV and SDP acknowledge financial support from the Spanish Ministerio de Econom\'{i}a
y Competitividad under grants AYA2016-79724-C4-4-P and PID2019-107408GB-C44, from Junta
de Andaluc\'{i}a Excellence Project P18-FR-2664, and also acknowledge support from the State Agency
  for Research of the Spanish MCIU through the `Center of Excellence Severo Ochoa' award for
  the Instituto de Astrof\'{i}sica de Andaluc\'{i}a (SEV-2017-0709). 
SDP is grateful to the Fonds de Recherche du Qu\'ebec - Nature et Technologies.  
The work is performed according to the Russian Government Program of Competitive Growth
of Kazan Federal University and Russian Science Foundation, grant no. 20-12-00105.  
This research has made use of the NASA/IPAC Extragalactic Database, which is funded by
the National Aeronautics and Space Administration and operated by the California Institute of Technology. 
We acknowledge the usage of the HyperLeda database (http://leda.univ-lyon1.fr). 
Funding for the Sloan Digital Sky Survey IV has been provided by the
Alfred P. Sloan Foundation, the U.S. Department of Energy Office of Science,
and the Participating Institutions. SDSS-IV acknowledges
support and resources from the Center for High-Performance Computing at
the University of Utah. The SDSS web site is www.sdss.org. 
SDSS-IV is managed by the Astrophysical Research Consortium for the
Participating Institutions of the SDSS Collaboration including the
Brazilian Participation Group, the Carnegie Institution for Science,
Carnegie Mellon University, the Chilean Participation Group,
the French Participation Group, Harvard-Smithsonian Center for Astrophysics,
Instituto de Astrof\'isica de Canarias, The Johns Hopkins University,
Kavli Institute for the Physics and Mathematics of the Universe (IPMU) /
University of Tokyo, Lawrence Berkeley National Laboratory,
Leibniz Institut f\"ur Astrophysik Potsdam (AIP), 
Max-Planck-Institut f\"ur Astronomie (MPIA Heidelberg),
Max-Planck-Institut f\"ur Astrophysik (MPA Garching),
Max-Planck-Institut f\"ur Extraterrestrische Physik (MPE),
National Astronomical Observatories of China, New Mexico State University,
New York University, University of Notre Dame,
Observat\'ario Nacional / MCTI, The Ohio State University,
Pennsylvania State University, Shanghai Astronomical Observatory,
United Kingdom Participation Group,
Universidad Nacional Aut\'onoma de M\'exico, University of Arizona,
University of Colorado Boulder, University of Oxford, University of Portsmouth,
University of Utah, University of Virginia, University of Washington, University of Wisconsin,
Vanderbilt University, and Yale University.


\newpage

\begin{appendix}

\section{Maps of the properties of galaxies using our measurements}

The figures in this section show the properties of galaxies derived from our measurements of the MaNGA spaxel spectra. 
In each figure, panel $a$ shows the distribution of the surface brightness in the photometric B-band across the image of the galaxy in sky coordinates (pixels).
North is up and east is left.  The pixel scale is 0.5 arcsec. The value of the surface brightness is color-coded.
The plus sign (or cross in some cases) shows the kinematic center of the galaxy, the dashed line is the major kinematic
axis, the dotted ellipse indicates the optical radius the galaxy. The solid ring marks the position of the spot with the enhanced
gas velocity dispersion. The size of the ring corresponds to the point spread function of the MaNGA measurements (2.5 arcsec or 5 pixels).
Panel $b$ shows the color-coded observed (line of sight) H${\alpha}$ velocity field of a given galaxy in sky coordinates.
Panel $c$ shows the color-coded oxygen abundance map. The oxygen abundances are determined through the R calibration from \citet{Pilyugin2016}.
Panel $d1$ shows the color-coded gas velocity dispersion $\sigma$ distribution across the image of the galaxy in sky coordinates. 
Panel $d2$ shows the  gas velocity dispersion as a function of radius for individual spaxels. The BPT type of the spectra is color-coded.
Panel $e$ shows the locations of the spaxels with spectra of different BPT types (the AGN-like,  H\,{\sc ii}-region-like,
and intermediate) on the image of the galaxy.
Panel $f1$ shows the examples of the spaxel spectra in the spot (blue line) and outside the spot (red line). The fluxes in the last spectrum are shifted 
for the sake of illustration. 
Panels $f2$ and $f3$ show the parts of the spectra from panel f1 around the H$\beta$ (panel f2) and H$\alpha$ (panel f3) lines.

\begin{figure*}
\begin{center}
\resizebox{1.00\hsize}{!}{\includegraphics[angle=000]{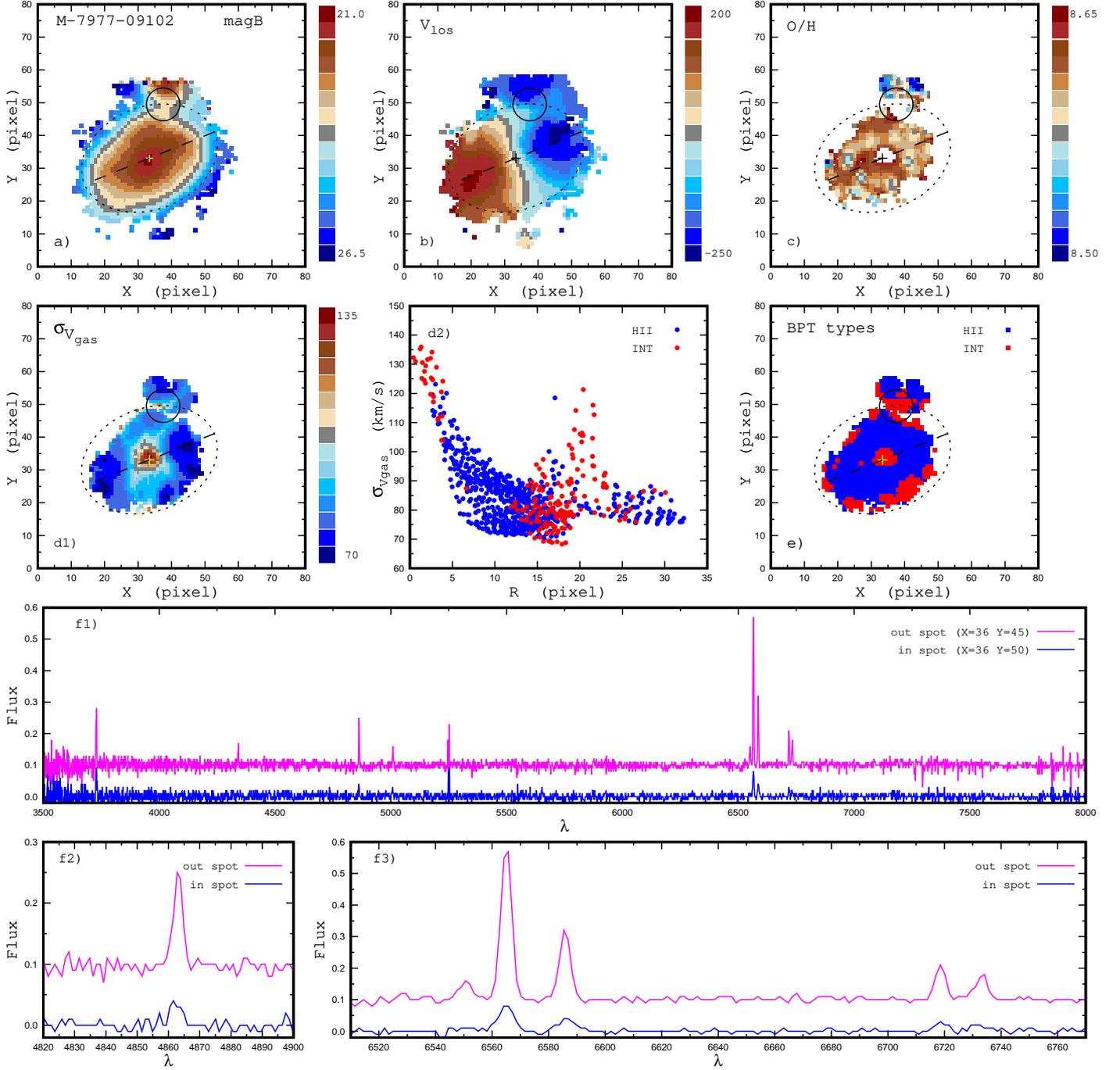}}
\caption{
  Properties of the MaNGA galaxy M-7977-09102. 
  {\em Panel} $a$:  distribution of the surface brightness in the photometric B-band across the image of the galaxy in sky coordinates (pixels). The value of the
  surface brightness is color-coded. The plus sign (or cross in some cases) shows the kinematic center of the galaxy, the dashed line is the major kinematic
  axis, the dotted ellipse indicates the optical radius the galaxy. The solid ring marks the position of the spot with the enhanced
  gas velocity dispersion. The size of the ring corresponds to the point spread function of the MaNGA measurements (2.5 arcsec or 5 pixel). 
  {\em Panel} $b$:  line-of-sight H${\alpha}$ velocity $V_{los}$ field.
  {\em Panel} $c$:  oxygen abundance map.
  {\em Panel} $d1$:  gas velocity dispersion $\sigma$ distribution across the image of the galaxy in sky coordinates. 
  {\em Panel} $d2$:  gas velocity dispersion as a function of radius for individual spaxels. The BPT type of the spectra is color-coded.
  {\em Panel} $e$: locations of the spaxels with spectra of different BPT types (the AGN-like,  H\,{\sc ii}-region-like,
  and intermediate) on the image of the galaxy.
  {\em Panel} $f1$: examples of the spaxel spectra in the spot (blue line) and outside the spot (red line). The fluxes in the last spectrum are shifted 
  for the sake of illustration. 
  {\em Panels} $f2$ and $f3$: parts of the spectra from panel f1 around the H$\beta$ (panel f2) and H$\alpha$ (panel f3) lines.
}
\label{figure:m-7977-09102}
\end{center}
\end{figure*}

\begin{figure*}
\begin{center}
\resizebox{1.00\hsize}{!}{\includegraphics[angle=000]{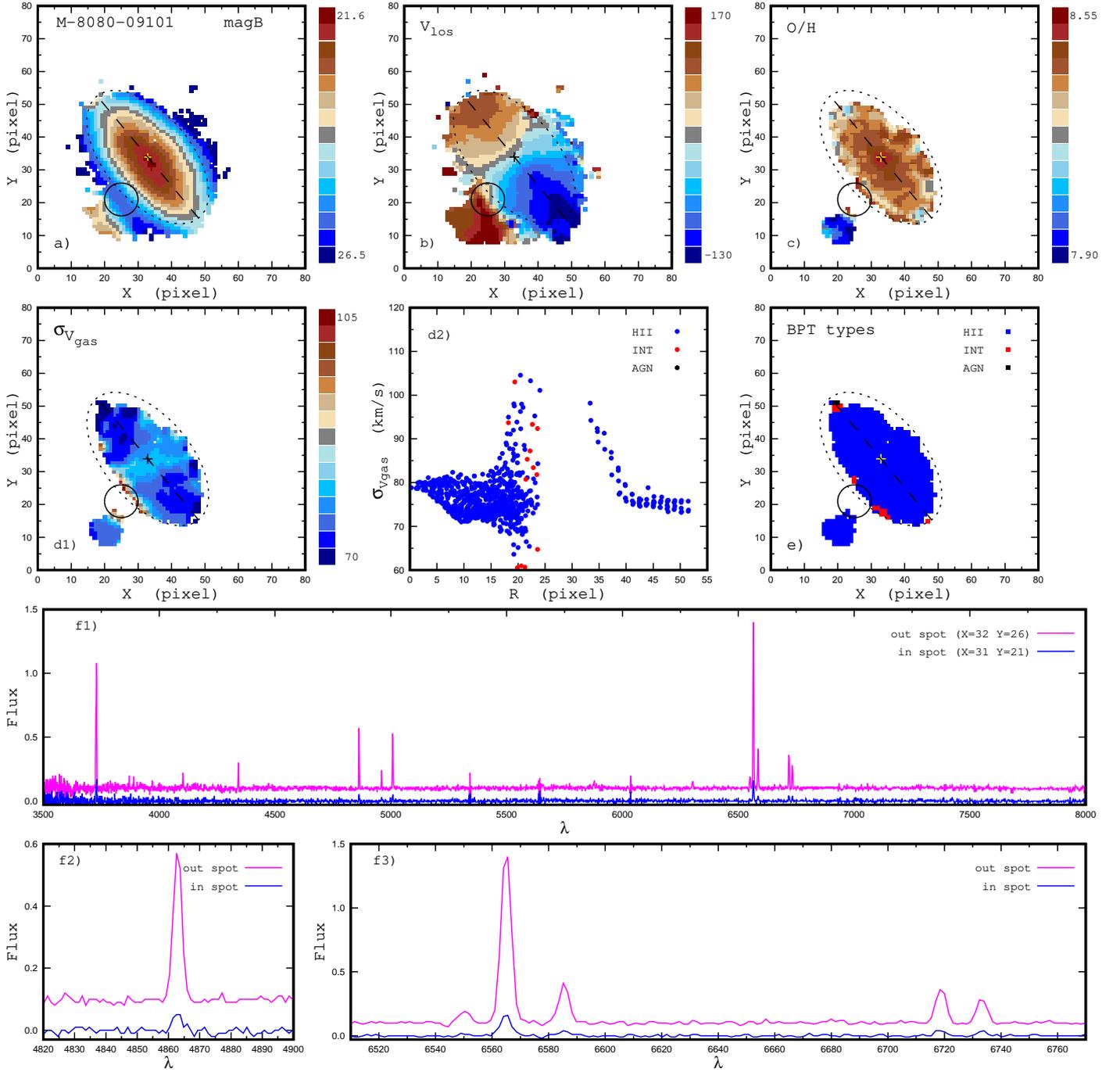}}
\caption{
  Properties of the MaNGA galaxy M-8080-09101. 
  The notation is the same as in Fig.~\ref{figure:m-7977-09102}.
}
\label{figure:m-8080-09101}
\end{center}
\end{figure*}

\begin{figure*}
\begin{center}
\resizebox{1.0\hsize}{!}{\includegraphics[angle=000]{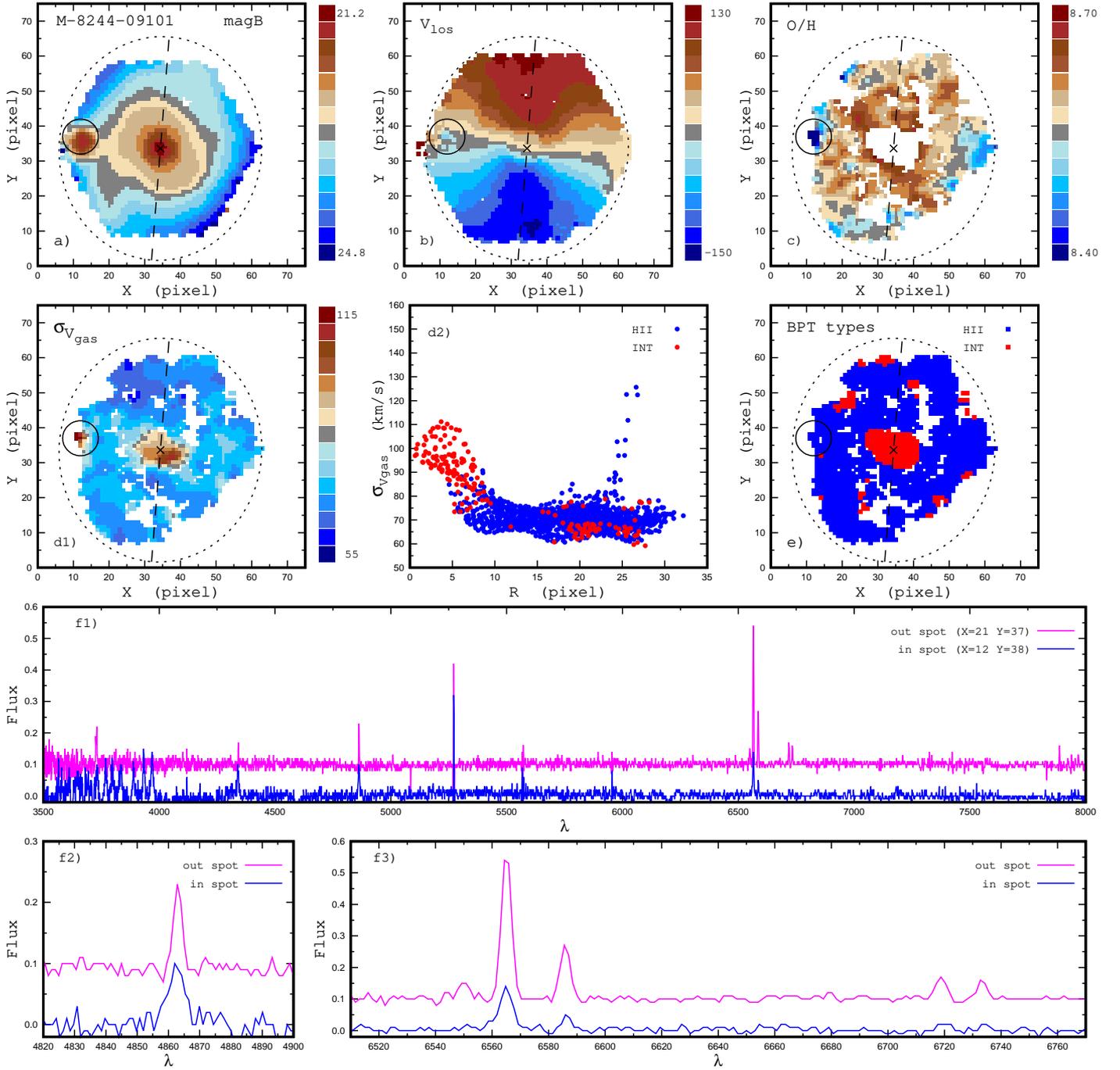}}
\caption{
  Properties of the MaNGA galaxy M-8244-09101 (PGC~025036). 
  The notation is the same as in Fig.~\ref{figure:m-7977-09102}.
}
\label{figure:m-8244-09101}
\end{center}
\end{figure*}

\begin{figure*}
\begin{center}
\resizebox{1.00\hsize}{!}{\includegraphics[angle=000]{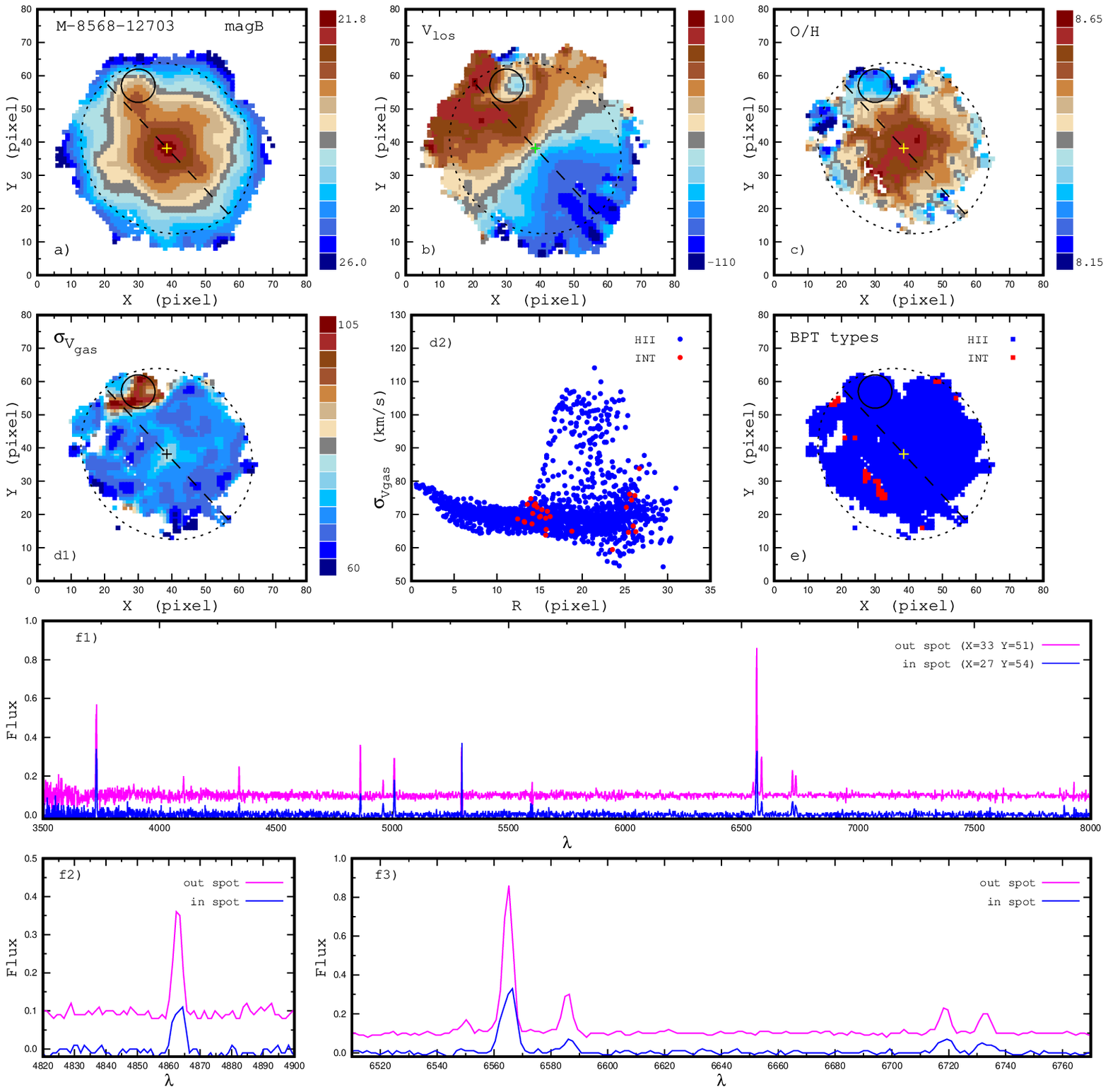}}
\caption{
  Properties of the MaNGA galaxy M-8568-12703. 
  The notation is the same as in Fig.~\ref{figure:m-7977-09102}.
}
\label{figure:m-8568-12703}
\end{center}
\end{figure*}

\begin{figure*}
\begin{center}
\resizebox{1.00\hsize}{!}{\includegraphics[angle=000]{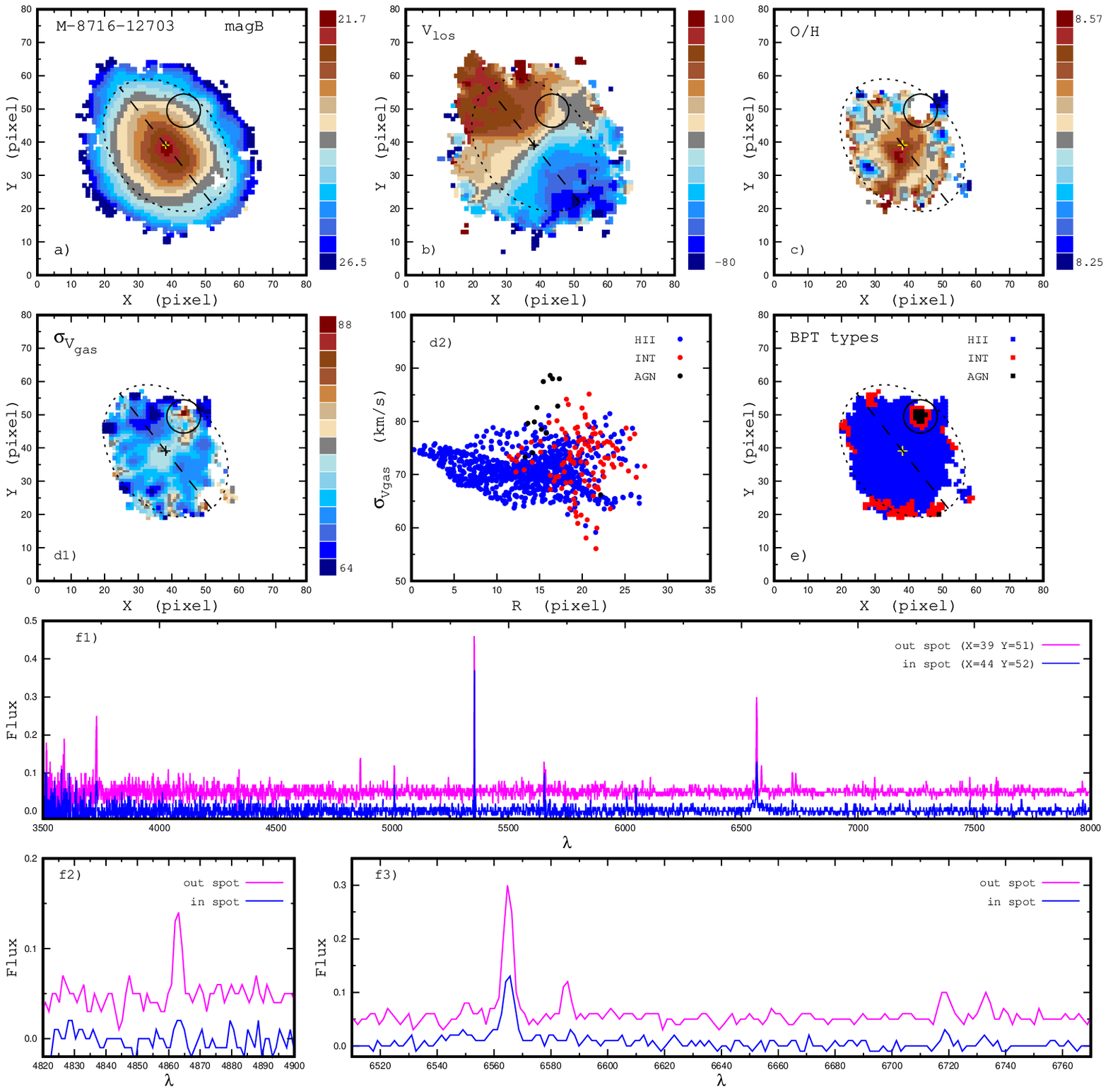}}
\caption{
  Properties of the MaNGA galaxy M-8716-12703. 
  The notation is the same as in Fig.~\ref{figure:m-7977-09102}.
}
\label{figure:m-8716-12703}
\end{center}
\end{figure*}

\begin{figure*}
\begin{center}
\resizebox{1.00\hsize}{!}{\includegraphics[angle=000]{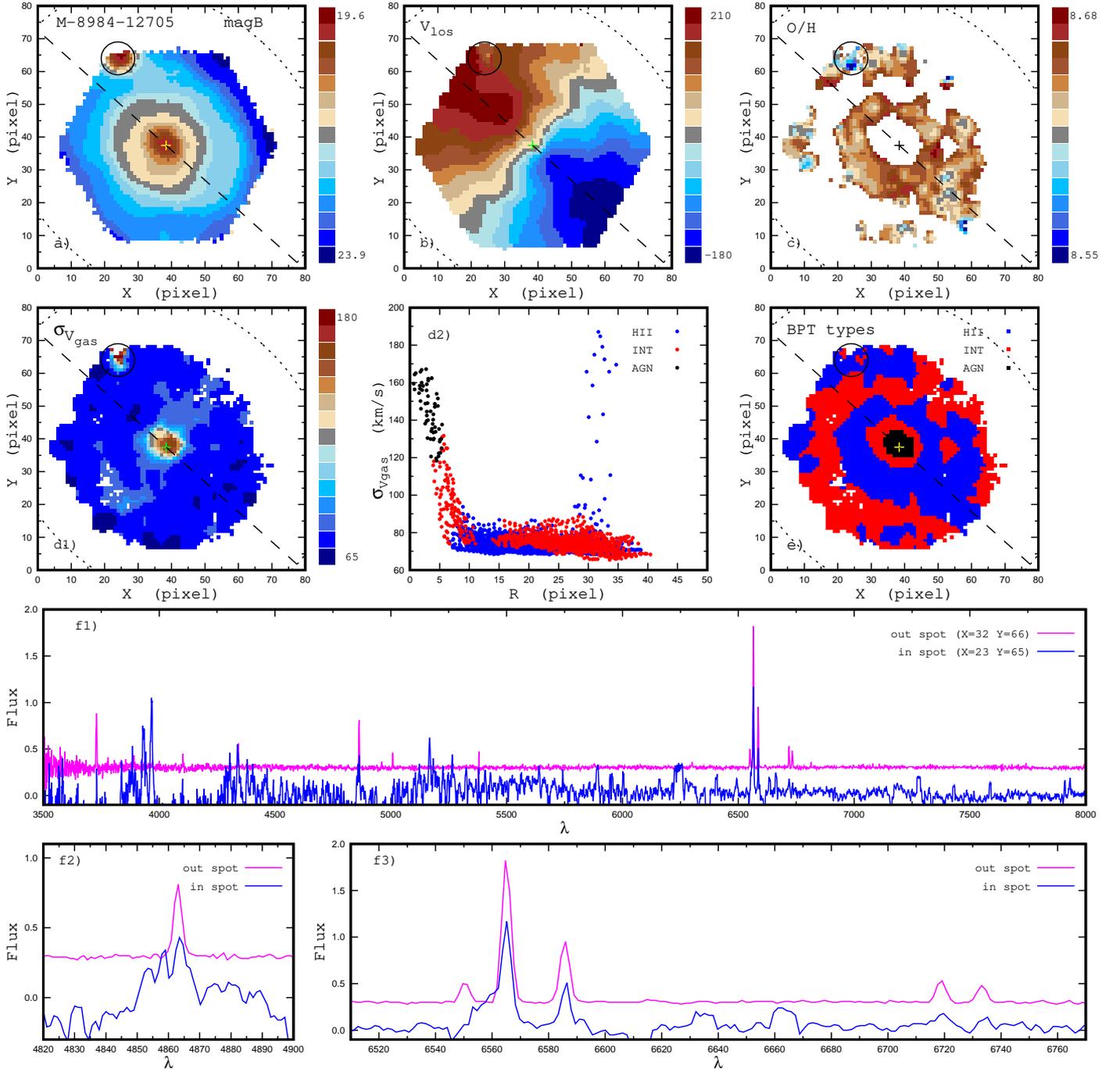}}
\caption{
  Properties of the MaNGA galaxy M-8984-12705 (NGC~5251). 
  The notation is the same as in Fig.~\ref{figure:m-7977-09102}.
}
\label{figure:m-8984-12705}
\end{center}
\end{figure*}

\begin{figure*}
\begin{center}
\resizebox{1.0\hsize}{!}{\includegraphics[angle=000]{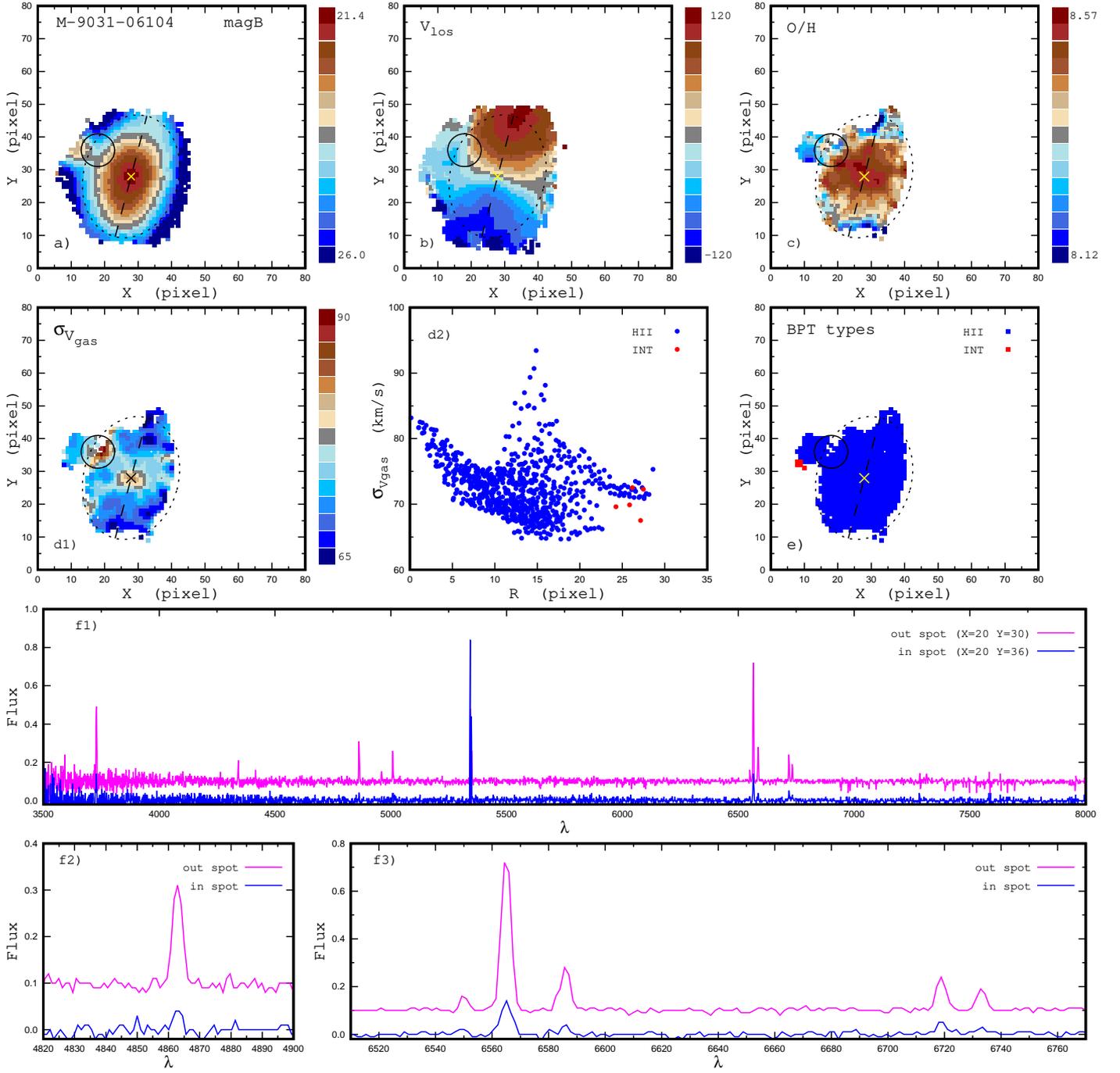}}
\caption{
  Properties of the MaNGA galaxy M-9031-06104. 
  The notation is the same as in Fig.~\ref{figure:m-7977-09102}.
}
\label{figure:m-9031-06104}
\end{center}
\end{figure*}

\begin{figure*}
\begin{center}
\resizebox{1.0\hsize}{!}{\includegraphics[angle=000]{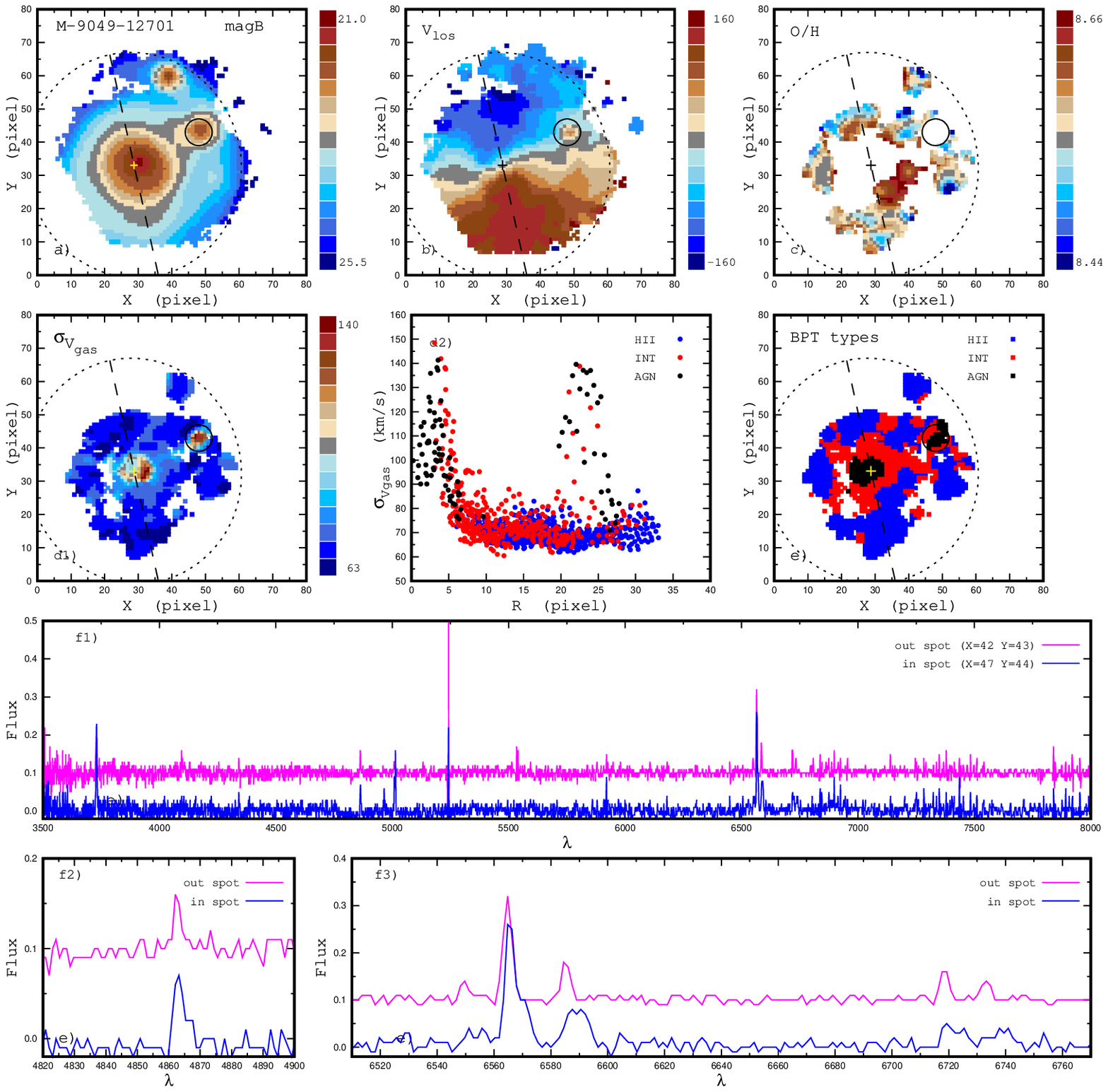}}
\caption{
  Properties of the MaNGA galaxy M-9049-12701 with double-peaked emission lines in the spot of the enhanced gas velocity dispersion. 
  The notation is the same as in Fig.~\ref{figure:m-7977-09102}.
}
\label{figure:m-9049-12701}
\end{center}
\end{figure*}

\section{Maps of the properties of galaxies for the MaNGA data analysis pipeline (DAP) measurements}

The figures in this section show the same as the figures in the previous section but the properties of galaxies are derived from the DAP measurements.

\begin{figure*}
\begin{center}
\resizebox{1.00\hsize}{!}{\includegraphics[angle=000]{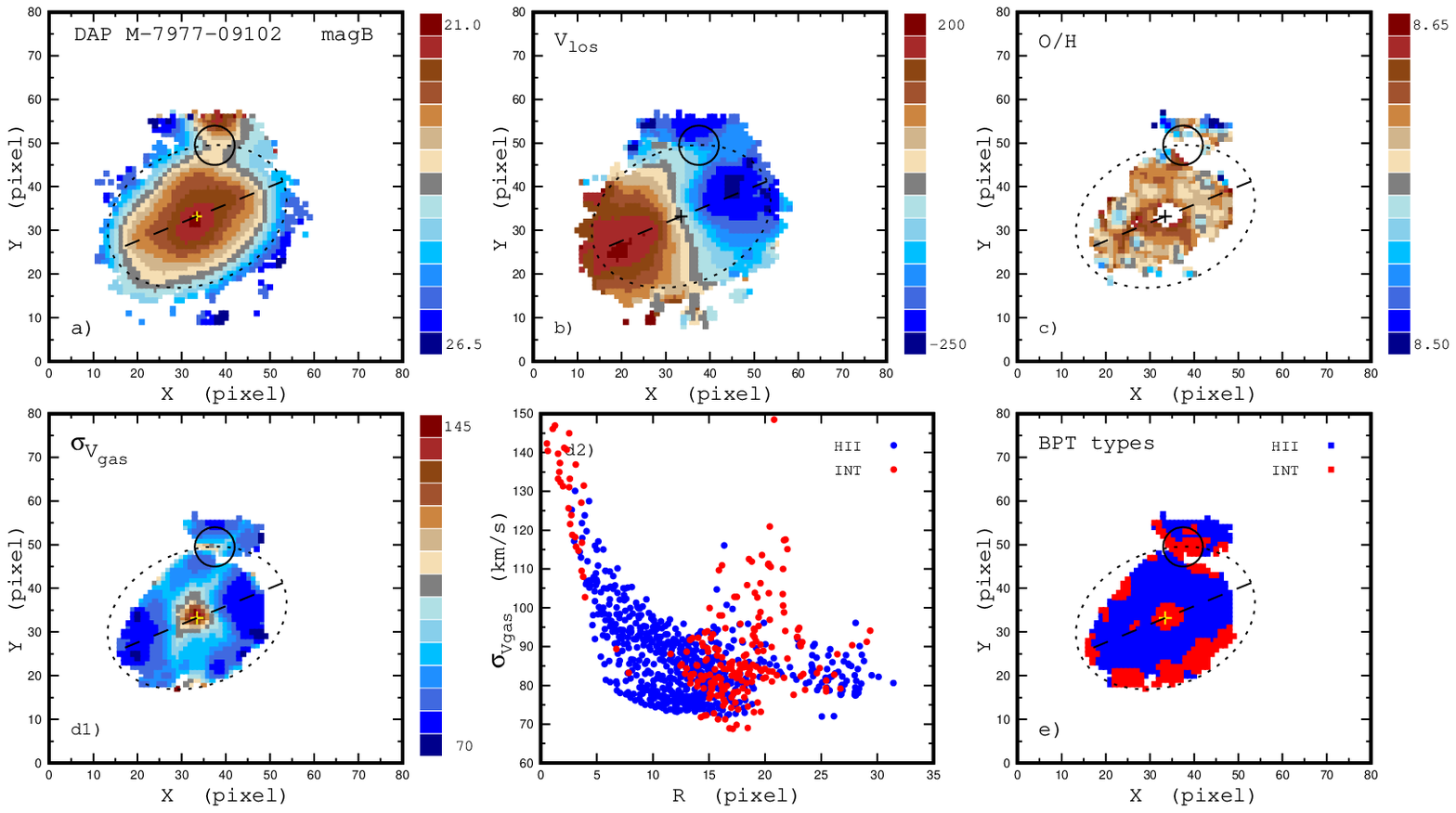}}
\caption{
  Properties of the MaNGA galaxy M-7977-09102 for the DAP measurements. 
  {\em Panel} $a$: distribution of the surface brightness in the photometric B-band across the image of the galaxy in sky coordinates (pixels). The value of the
  surface brightness is color-coded. The plus sign (or cross in some cases) shows the kinematic center of the galaxy, the dashed line is the major kinematic
  axis, the dotted ellipse indicates the optical radius the galaxy. The solid ring marks the position of the spot with the enhanced
  gas velocity dispersion. The size of the ring corresponds to the point spread function of the MaNGA measurements (2.5 arcsec or 5 pixel). 
  {\em Panel} $b$:  line-of-sight H${\alpha}$ velocity $V_{los}$ field.
  {\em Panel} $c$:  oxygen abundance map.
  {\em Panel} $d1$:  gas velocity dispersion $\sigma$ distribution across the image of the galaxy in sky coordinates. 
  {\em Panel} $d2$:  gas velocity dispersion as a function of radius for individual spaxels. The BPT type of the spectra is color-coded.
  {\em Panel} $e$: locations of the spaxels with spectra of different BPT types (the AGN-like,  H\,{\sc ii}-region-like,
  and intermediate) on the image of the galaxy.
}
\label{figure:m-7977-09102-dap}
\end{center}
\end{figure*}

\begin{figure*}
\begin{center}
\resizebox{1.00\hsize}{!}{\includegraphics[angle=000]{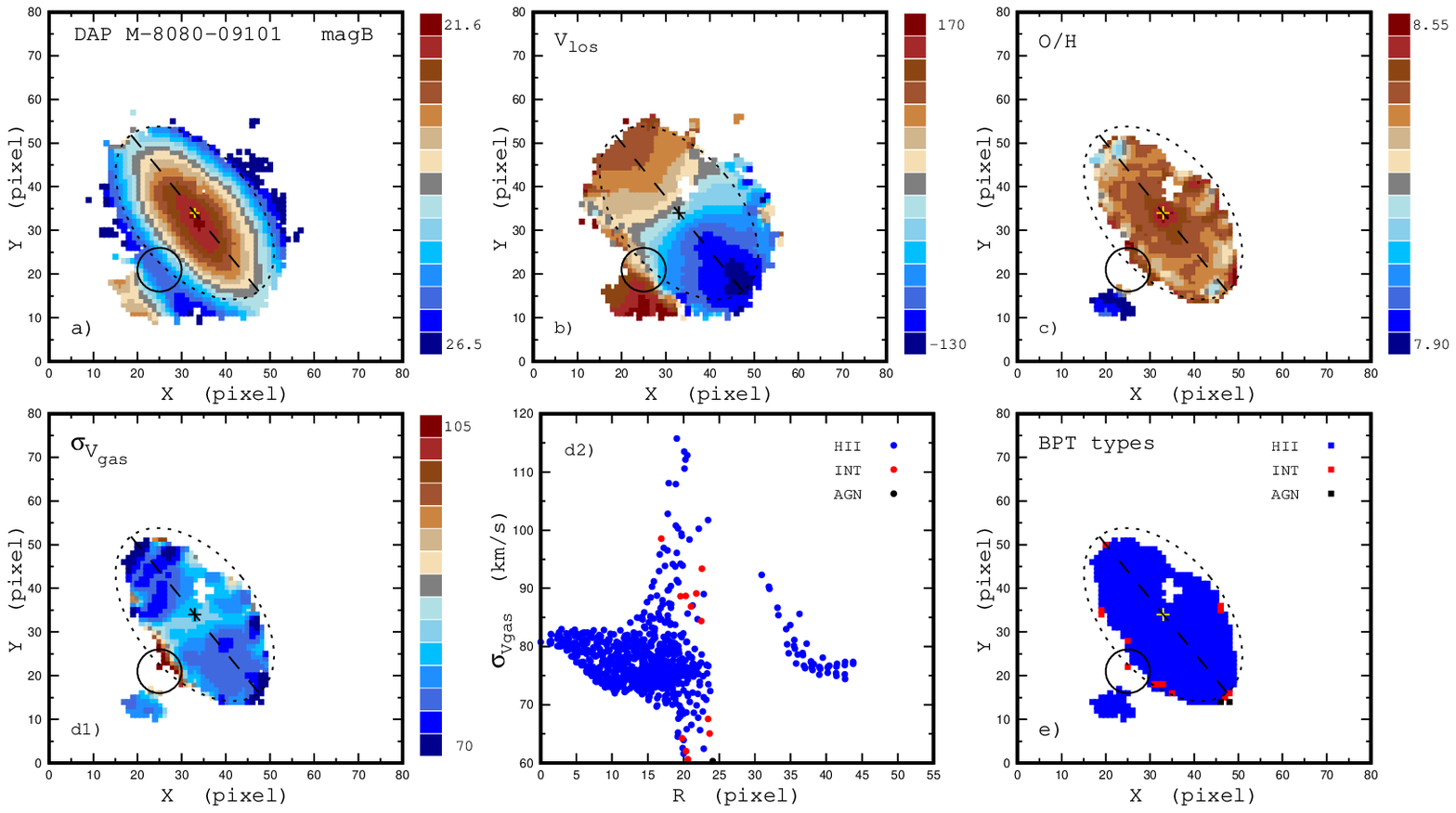}}
\caption{
  Properties of the MaNGA galaxy M-8080-09101 for the DAP measurements.  
  The notation is the same as in Fig.~\ref{figure:m-7977-09102-dap}.
}
\label{figure:m-8080-09101-dap}
\end{center}
\end{figure*}

\begin{figure*}
\begin{center}
\resizebox{1.0\hsize}{!}{\includegraphics[angle=000]{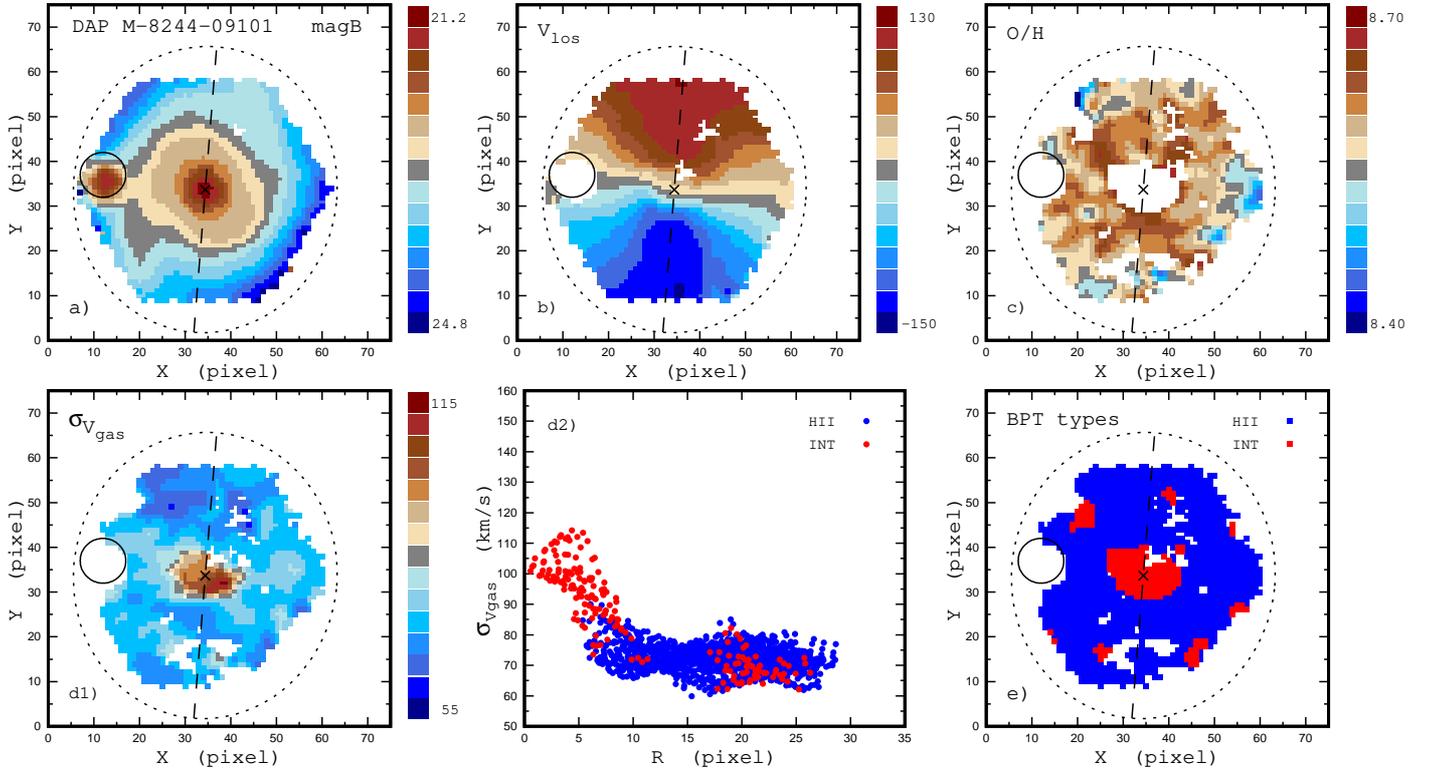}}
\caption{
  Properties of the MaNGA galaxy M-8244-09101 (PGC~025036) for the DAP measurements. 
  The notation is the same as in Fig.~\ref{figure:m-7977-09102-dap}.
}
\label{figure:m-8244-09101-dap}
\end{center}
\end{figure*}

\begin{figure*}
\begin{center}
\resizebox{1.00\hsize}{!}{\includegraphics[angle=000]{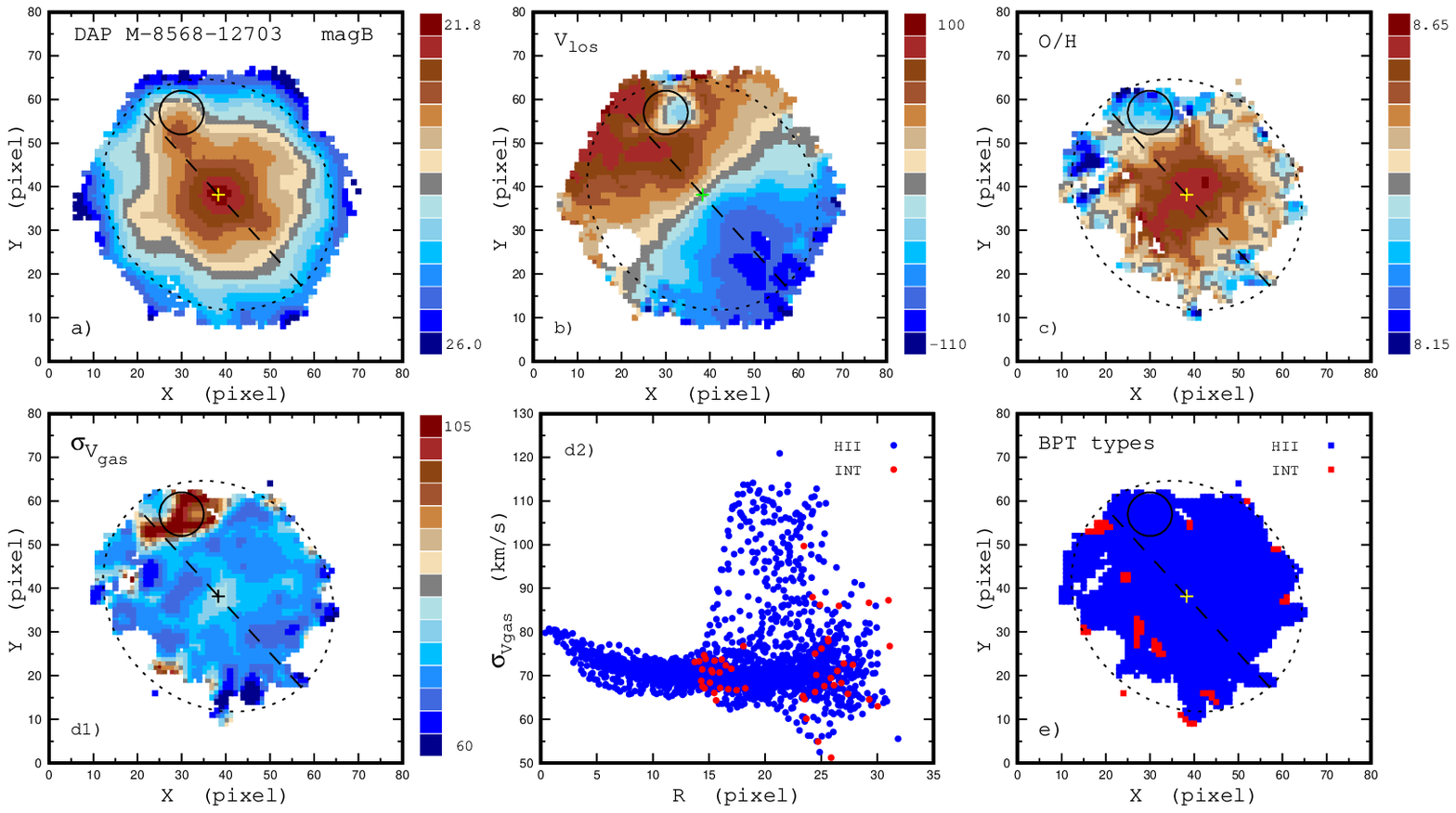}}
\caption{
  Properties of the MaNGA galaxy M-8568-12703 for the DAP measurements. 
  The notation is the same as in Fig.~\ref{figure:m-7977-09102-dap}.
}
\label{figure:m-8568-12703-dap}
\end{center}
\end{figure*}

\begin{figure*}
\begin{center}
\resizebox{1.00\hsize}{!}{\includegraphics[angle=000]{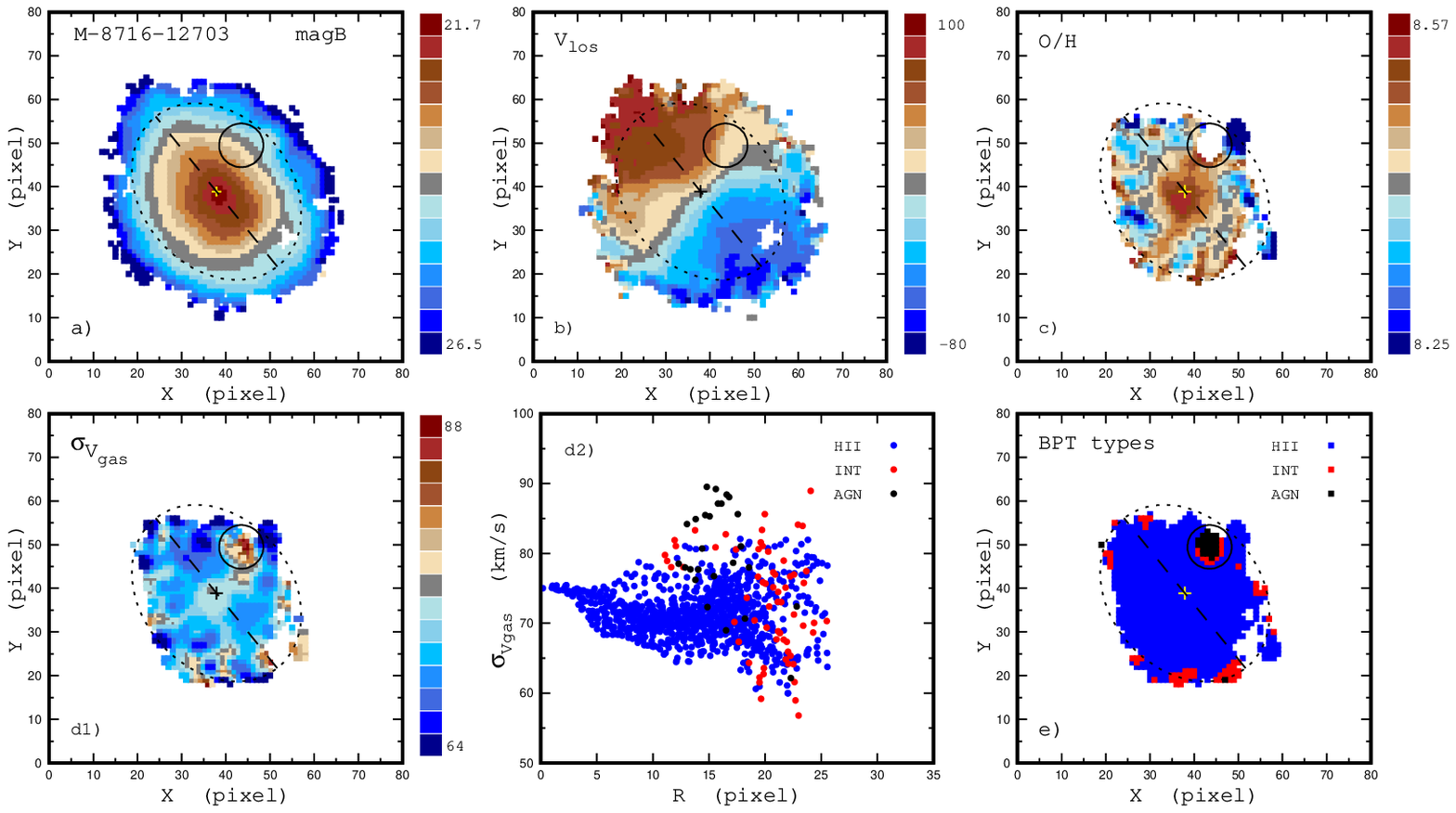}}
\caption{
  Properties of the MaNGA galaxy M-8716-12703 for the DAP measurements. 
  The notation is the same as in Fig.~\ref{figure:m-7977-09102-dap}.
}
\label{figure:m-8716-12703-dap}
\end{center}
\end{figure*}




\begin{figure*}
\begin{center}
\resizebox{1.00\hsize}{!}{\includegraphics[angle=000]{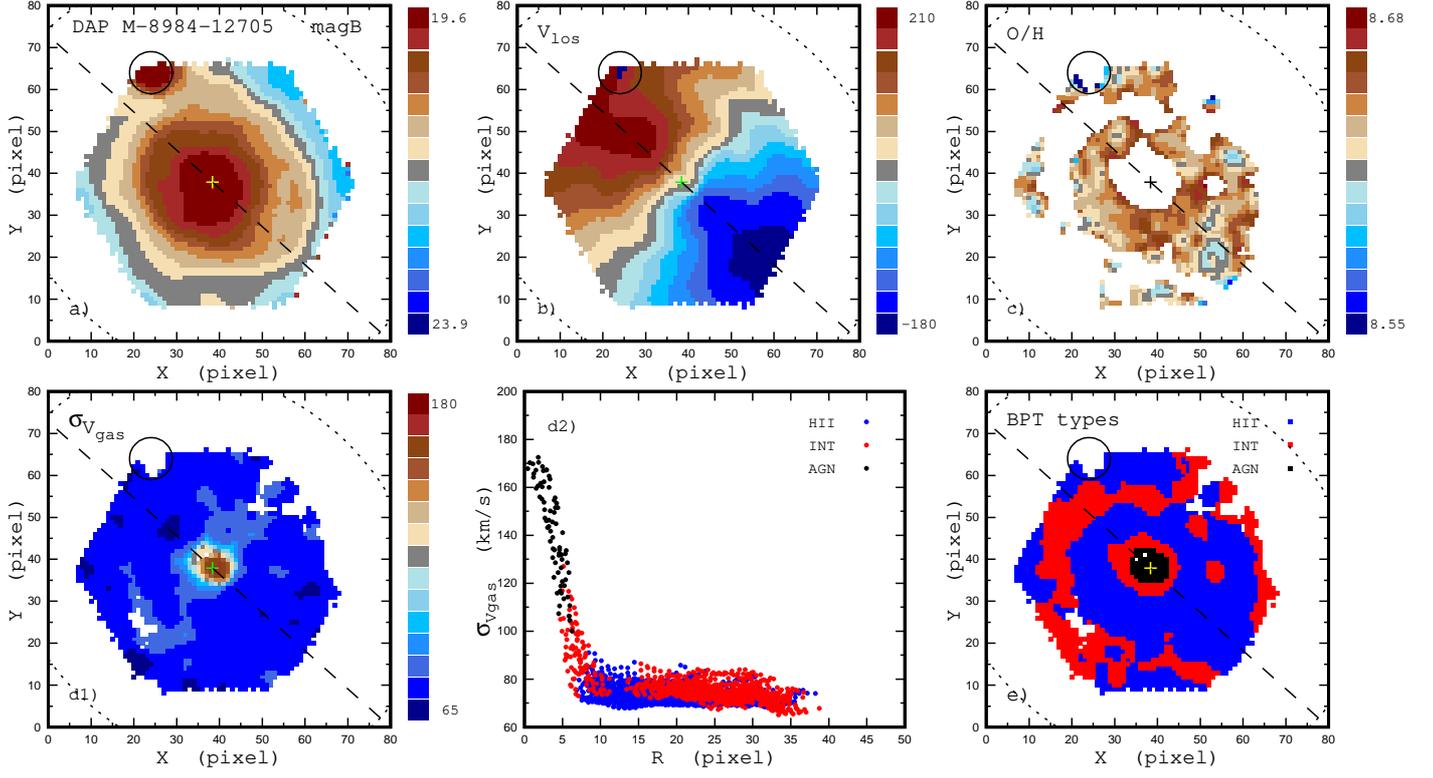}}
\caption{
  Properties of the MaNGA galaxy M-8984-12705 (NGC~5251) for the DAP measurements. 
  The notation is the same as in Fig.~\ref{figure:m-7977-09102-dap}.
}
\label{figure:m-8984-12705-dap}
\end{center}
\end{figure*}

\begin{figure*}
\begin{center}
\resizebox{1.0\hsize}{!}{\includegraphics[angle=000]{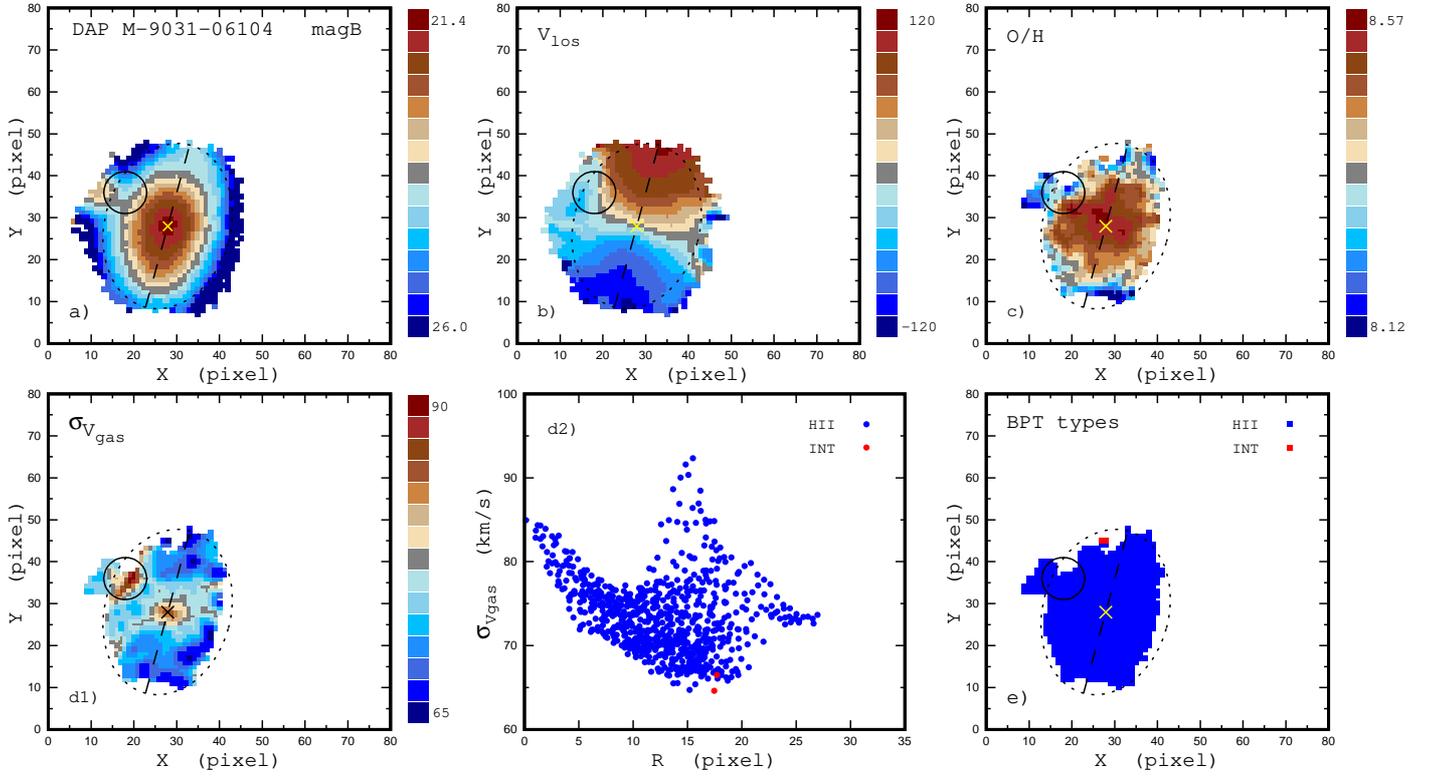}}
\caption{
  Properties of the MaNGA galaxy M-9031-06104 for the DAP measurements. 
  The notation is the same as in Fig.~\ref{figure:m-7977-09102-dap}.
}
\label{figure:m-9031-06104-dap}
\end{center}
\end{figure*}

\end{appendix}

\end{document}